\documentclass[useAMS,usenatbib]{mn2e}

\usepackage{aas_macros}
\usepackage{amsmath}
\usepackage{amssymb}
\usepackage{graphicx,graphics}
\usepackage{rotating}
\usepackage{natbib}

\usepackage{times}

\usepackage{hyperref}
\hypersetup
{
    pdfauthor={Kokubo},
    pdftitle={kokubo2014},
}
\title[Constraints on the quasar inhomogeneous disc model]{Constraints on the temperature inhomogeneity in quasar accretion discs
from the ultraviolet-optical spectral variability}
\author[Mitsuru Kokubo]{Mitsuru Kokubo$^{1,2}$
\thanks{E-mail: mkokubo@ioa.s.u-tokyo.ac.jp}\\
$^{1}$Department of Astronomy, School of Science, the University of Tokyo, 7-3-1 Hongo, Bunkyo-ku, Tokyo 113-0033, Japan\\
$^{2}$Institute of Astronomy, the University of Tokyo, 2-21-1 Osawa, Mitaka, Tokyo 181-0015, Japan}
\begin{document}

\date{Accepted for publication in Monthly Notices of the Royal Astronomical Society 2015 February 04}

\pagerange{\pageref{firstpage}--\pageref{lastpage}} \pubyear{2015}

\maketitle

\label{firstpage}

\begin{abstract}

The physical mechanisms of the quasar ultraviolet (UV)-optical variability are not well understood despite the long history of observations. 
Recently, Dexter \& Agol presented a model of quasar UV-optical variability, which assumes large local temperature fluctuations in the quasar accretion discs.
This inhomogeneous accretion disc model is claimed to describe not only the single-band variability amplitude, but also microlensing size constraints and the quasar composite spectral shape.
In this work, we examine the validity of the inhomogeneous accretion disc model in the light of quasar UV-optical spectral variability by using five-band multi-epoch light curves for nearly 9 000 quasars in the Sloan Digital Sky Survey (SDSS) Stripe 82 region.
By comparing the values of the intrinsic scatter $\sigma_{\text{int}}$ of the two-band magnitude$-$magnitude plots for the SDSS quasar light curves and for the simulated light curves, we show that Dexter \& Agol's inhomogeneous accretion disc model cannot explain the tight inter-band correlation often observed in the SDSS quasar light curves. 
This result leads us to conclude that the local temperature fluctuations in the accretion discs are not the main driver of the several years' UV-optical variability of quasars, and consequently, that the assumption that the quasar accretion discs have large localized temperature fluctuations is not preferred from the viewpoint of the UV-optical
spectral variability.

\end{abstract}


\section{Introduction}
\label{intro}

The flux variability of active galactic nuclei (AGN) and quasars has been observed for a long time since its discovery. 
The variability amplitude in the ultraviolet (UV)-optical wavelength range reaches almost an order of magnitude over several years, which indicates that the mechanism causing the variability in an AGN is the mechanism controlling the AGN activity \citep[e.g.,][]{ulr97, nan98, kaw98, giv99, haw02, van04, dev05, gas08, wil08, liu08, bau09, sch12, meu13, kok14, rua14}.

There are a number of models attempting to account for the AGN UV-optical variability. 
Several models assume that the AGN variability is caused by external (non-AGN) factors, e.g., gravitational microlensing \citep{haw93,haw02}, star collisions \citep{tor00} or multiple supernovae or starbursts near the nucleus \citep{tel92,are94,are97,cid00}.
However, these models generally fail to account for the ubiquity of AGN variability that has been confirmed by modern large time-domain surveys \citep[e.g.,][]{van04}.
Alternatively, several authors have claimed that the AGN variability is due to changes in the global mass accretion rate in AGN accretion discs \citep{per06, li08, sak11, zuo12, gu13a}.
The variable mass accretion rate model seems to explain the large variability amplitude and the bluer-when-brighter colour variability trend often observed for AGNs \cite[e.g.,][]{cut85,dic96,cri97,ulr97,giv99,haw03,van04,meu11,sak11,zuo12,sch12,gez13,kok14,gal14}. 
However, as shown by \cite{kok14} \citep[see also][]{sch12}, the variable mass accretion rate model cannot fully account for the strong variability observed in UV wavelengths. 
Moreover, the variable mass accretion rate model is not preferred when the large difference between the AGN UV-optical spectral variability timescale and the sound crossing (and the viscous) timescale of the accretion disc, which corresponds to the timescale required for global changes of mass accretion rate within the whole of the accretion disc, is taken into consideration  \citep[see e.g.,][]{cou91,lam14,utt14}.

Recently, an alternative AGN variability model was presented by \cite{dex11}: the strongly inhomogeneous accretion disc model, which assumes large local temperature fluctuations in the quasar accretion discs as the cause of flux variability. 
Dexter \& Agol's inhomogeneous accretion disc model is motivated by recent numerical simulation studies of thermal or magnetorotational instabilities \citep[e.g.,][and references therein]{hir09a,jiang13}.
This model aims to describe not only the observed amplitude of AGN UV-optical single-band variability, but also the unexpectedly large size of the quasar accretion disc revealed by microlensing observations \citep[see e.g.,][]{poo07,dai10,jim14,ede15}, the excess emission in UV wavelengths observed in a composite Hubble Space Telescope spectrum of quasars \citep{zhe97,kaw01}, and the stochastic properties of quasar light curves \citep[][and references therein]{kel09, koz10, mac12, zu13, and13, mor14}.
The Dexter \& Agol inhomogeneous accretion disc model has received much attention, and several authors have claimed that this model can reproduce several observed properties of quasar UV-optical variability \citep[e.g.,][]{meu13,rua14,sun14}.
For example, \cite{meu13} discussed that the observed anti-correlation between the variability amplitude and the mass accretion rate of quasars could be explained by the Dexter \& Agol inhomogeneous accretion disc model. 
\cite{sun14} discussed that the Dexter \& Agol inhomogeneous accretion disc model seemed (at least qualitatively) to account for the
timescale-dependent colour variability of quasars.

\cite{rua14} constructed a composite difference spectrum using two-epoch quasar spectra (for 604 quasars) from the Sloan Digital Sky Survey (SDSS) \citep{yor00} following the procedure of \cite{wil05}, and compared it with the Dexter \& Agol inhomogeneous accretion disc model. 
The model's composite difference spectra are calculated as follows: 5 000 model difference spectra between any two successive time steps are produced, and then the model's composite difference spectrum is produced by taking a geometric mean. 
In \cite{rua14}, they obtained a reasonable fitting of the model's geometric mean composite difference spectra to the observed composite difference spectrum, and concluded that quasar UV-optical variability is mainly caused by the large localized temperature fluctuations in the quasar accretion discs.

However, it should be noted that because the quasar accretion discs have a large physical size, temperature fluctuations occurring at different radii of the quasar accretion discs, which have different mean temperatures, must be causally unconnected. 
Thus, the superposition of independent flares from the localized temperature fluctuations generally means that the inter-band flux$-$flux correlation observed in the two-band light curve of each individual quasar becomes weaker and weaker when we take more and more separated wavelength band pairs, which contradicts a well-known feature of AGN variability: the UV-optical continua light curves of quasars in different bands are highly correlated \citep[e.g.,][]{cho81, kro91, cou91, kor95, ulr97, win97, cac07, gas08, sak10, kok14}.
According to this consideration, the successful fitting to the observed difference spectrum by the Dexter \& Agol inhomogeneous accretion disc model obtained by \cite{rua14} is expected to be obtainable only when the comparison between the model and observed difference spectra is done in a composite sense, because the model's predicted weak inter-band correlation can be smeared out when compositing the model spectra. 
As mentioned in \cite{kok14}, at least qualitatively, it seems to be difficult to explain the large coherent inter-band variation within the UV-optical wavelength range (i.e., the strong flux$-$flux correlation for each individual quasar) by localized flares in the quasar accretion discs.

In this paper, we give quantitative counterarguments against Dexter \& Agol's inhomogeneous accretion disc model from the viewpoint of quasar UV-optical spectral variability. 
We calculate the intrinsic scatter (see Section~\ref{method} for details) of the inter-band linear correlation in magnitude$-$magnitude space predicted by the Dexter \& Agol inhomogeneous accretion disc model and compare it with the observed scatter quantitatively. 
As a result of the comparison, we show that the Dexter \& Agol inhomogeneous accretion disc model actually cannot explain the tight magnitude$-$magnitude correlation often observed in quasar UV-optical multi-band light curves. 
We conclude that it is suspicious that the large localized temperature fluctuations are the main driver of the quasar variability, and consequently, that the assumption of a strongly inhomogeneous accretion disc is not preferred from the viewpoint of the UV-optical spectral variability.

In Section \ref{method}, we introduce the statistical method we use in this work to evaluate the intrinsic scatter of the magnitude$-$magnitude plots of the two-band light curves. 
We describe the database of the SDSS Stripe~82 multi-band multi-epoch light curves of quasars, and then evaluate the variability amplitude and the intrinsic scatter of the magnitude$-$magnitude plots of these light curves in Section~\ref{data}.
We describe the Dexter \& Agol inhomogeneous accretion disc model
in Section~\ref{formalization}, and then show the details of the model light curve calculations in Section~\ref{simulation}. 
We compare the data with the model predictions in Section~\ref{model_comparison}. 
Finally, discussion and conclusions are given in Section~\ref{discussion}.

\section{Method}
\label{method}

\begin{figure}
\center{
\includegraphics[clip, width=6.2in, angle=90]{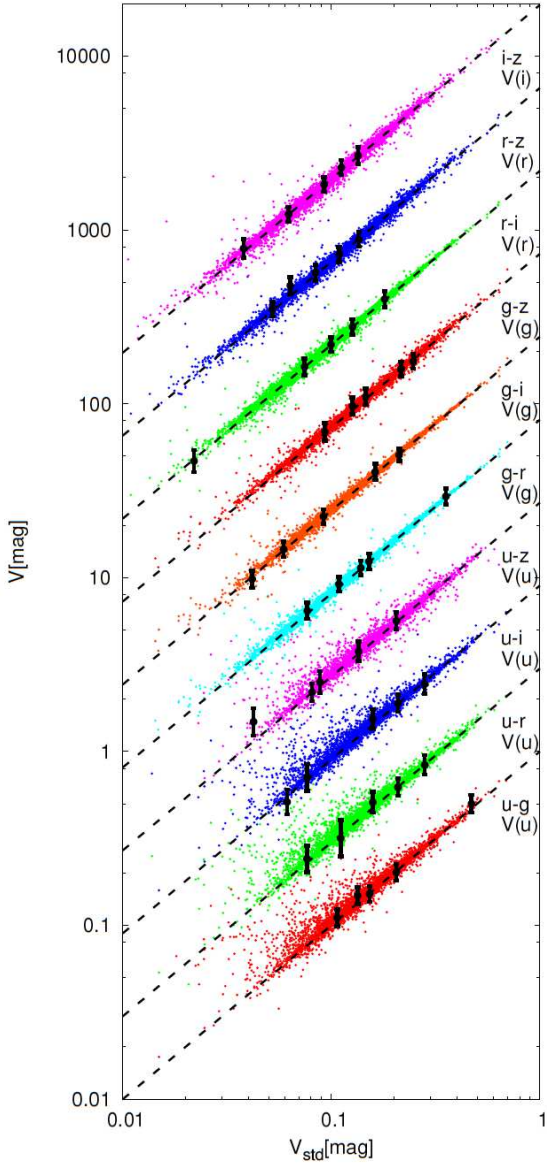}
}
\caption{Comparison of the values of $u$-, $g$-, $r$-, and $i$-band variability amplitudes for the SDSS Stripe~82 quasars derived by the two methods.
$V_{\text{std}}$ is defined as $V_{\text{std}}=\sqrt{\Sigma^2 - \xi^2}$, where $\Sigma$ is the sample standard deviation and $\xi$ is the root mean square of the photometric error of a single-band light curve.
$V$ is the point estimate value of the variability amplitude derived by the Bayesian regression analyses from {\tt LINMIX\_ERR} for the magnitude$-$magnitude plots of the band pairs of $u$-$g$, $u$-$r$, $u$-$i$, $u$-$z$ for $V$($u$), $g$-$r$, $g$-$i$, $g$-$z$ for $V$($g$), $r$-$i$, $r$-$z$ for $V$($r$), and $i$-$z$ for $V$($i$).
The values of $V$ for adjacent band pairs are scaled by 3 for clarity.
Black points with error bars are the randomly selected measurement points indicating the typical values of uncertainty in $V$, and straight lines indicate $V = V_{\text{std}}$.
}
\label{fig:consistency}
\end{figure}

The flux$-$flux plots, and almost equivalently the magnitude$-$magnitude plots, of quasar two-band simultaneous light curves are known to be well fitted by straight lines of $y = a + bx$ \citep[e.g.,][]{cho81,win92,hag97,cac07,sak10,sak11,lir11,sch12,kok14}.
The linear regression slope of the flux$-$flux plot of an AGN light curve quantifies the colour of the variable component \citep[][]{hag97,kok14}, and the linear regression slope of the magnitude$-$magnitude plot can be an indicator of the colour variability of AGNs \citep[][]{sch12}.
In this work, as described later, we focus on the intrinsic scatter from the regression line in magnitude$-$magnitude space, which can be considered as an indicator of the strength of the inter-band correlation; the larger the intrinsic scatter is, the weaker the two-band magnitude$-$magnitude correlation is. 
Since the values of the variability amplitude and the intrinsic scatter evaluated in magnitude$-$magnitude space do not depend on the absolute value of the flux (i.e., the magnitude unit is dimensionless), it is preferable to use the magnitude$-$magnitude plot because it makes the comparison with model predictions easy. 
Thus, in later sections we use the magnitude unit when quantifying the quasar variability.

The linear regression line of a magnitude$-$magnitude plot of a two-band quasar light curve with measurement errors on both axes and with intrinsic scatter can be modelled as [following the notation of \citealt{kel07}]
\begin{eqnarray}
\eta_i &=& a + b\xi_i + \epsilon_i\\
x_i &=& \xi_i  + \epsilon_{x,i}\\
y_i &=& \eta_i + \epsilon_{y,i}\\
\epsilon_i &\sim & G\left(\sigma^2_{\text{int}}\right),\ \epsilon_{x,i}\sim G\left(\sigma_{x,i}^2\right),\ \epsilon_{y,i}\sim G\left(\sigma_{y,i}^2\right),
\end{eqnarray}
where the measurement values of the two-band magnitudes are expressed as $x_i$ and $y_i$, in which $i$ labels the measurement epoch; $i=$1, 2, $\cdots$.
$G(x^2)$ indicates a Gaussian distribution function with the variance $x^2$ and with zero mean, and a tilde ($\sim$) means that a variable on the left side is drawn from a distribution function on the right side.
$\xi_i$ is the independent variable and $\eta_i$ is the dependent variable, representing the true values of the two-band magnitudes.
$\epsilon_{x,i}$ and $\epsilon_{y,i}$ are the random measurement errors on $x_i$ and $y_i$, whose variances are $\sigma_{x,i}^2$ and $\sigma_{y,i}^2$, respectively.
$\epsilon_i$ represents the intrinsic scatter, whose variance is assumed to be constant and is denoted as $\sigma^2_{\text{int}}$, which corresponds to the square of the standard error of the regression.

\begin{table*}
\caption{Properties of the individual filter combinations for the SDSS Stripe 82 quasar light curves.}
\label{tbl:data_properties}
\begin{tabular}{@{}cccccccc}
\hline
Band pairs ($s$-$l$) & $z$ & $\langle z \rangle$ & $N_{\text{obj}}$ & $\langle$ $N_{\text{epoch}}$ $\rangle$ & $\langle \lambda_{\text{rest}} \rangle_s$ & $\langle \lambda_{\text{rest}} \rangle_l$& $\Delta \langle \lambda_{\text{rest}} \rangle$ \\
\hline
$u$-$g$&   $0.0-1.5$&$1.025$&$3984$&$53$&$1754$&$2314$&$560$ \\
$u$-$r$&   $0.0-1.5$&$1.025$&$3977$&$54$&$1754$&$3044$&$1290$ \\
$u$-$i$&   $0.0-1.5$&$1.026$&$3964$&$54$&$1753$&$3692$&$1939$ \\
$u$-$z$&   $0.0-1.5$&$1.010$&$3502$&$52$&$1767$&$4443$&$2676$ \\
$g$-$r$&   $0.0-2.0$&$1.280$&$6715$&$58$&$2055$&$2704$&$649$ \\
$g$-$i$&   $0.0-2.0$&$1.280$&$6708$&$58$&$2055$&$3281$&$1226$ \\
$g$-$z$&   $0.0-2.0$&$1.263$&$5787$&$56$&$2071$&$3947$&$1876$ \\
$r$-$i$&   $0.0-3.5$&$1.558$&$8783$&$60$&$2410$&$2925$&$515$ \\
$r$-$z$&   $0.0-3.5$&$1.517$&$7417$&$57$&$2449$&$3548$&$1099$ \\
$i$-$z$&   $0.0-4.5$&$1.547$&$7510$&$57$&$2937$&$3506$&$569$ \\\hline
\end{tabular}

\medskip 
A band pair is expressed as $s$-$l$, where $s$ is the band with the $shorter$ effective wavelength and $l$ indicates the other one. $\langle z \rangle$, $N_{\text{obj}}$, and $\langle N_{\text{epoch}} \rangle$ are the mean redshift, the number of objects and the mean number of observed epochs for the SDSS Stripe~82 light curve sample of each band pair (see Section~\ref{data} for the details of the sample selection). $\langle \lambda_{\text{rest}} \rangle_s$ and $\langle \lambda_{\text{rest}} \rangle_l$ are the mean rest-frame wavelengths of the bands, defined as the effective wavelengths of the bands divided by $1+\langle z \rangle$. $\Delta \langle \lambda_{\text{rest}} \rangle$ is their difference (in units of \AA).
\end{table*}

There are several methods for evaluating the linear regression
line and the intrinsic scatter for data with measurement errors on
both axes \citep[][and references therein]{gul89,dag05,kel07,hog10,wal12,fei12,par12}.
\cite{kel07} introduced a Bayesian approach to this problem.
\cite{kel07}'s method assumes that the measurement errors and intrinsic scatter are Gaussian, and the probability distribution of independent variable $\xi_{i}$ (true value of the $i$th measurement $x_{i}$), denoted as $p(\xi_{i}|\psi)$ where $\psi$ is the parameter set describing the distribution, is modelled as a weighted mixture of $K$ Gaussian functions. 
\cite{kel07} mentioned that it was useful to model $p(\xi_{i}|\psi)$ using this form because it was flexible enough to adapt to a wide variety of distributions. 
In the particular case of the magnitude$-$magnitude plots for the quasar light curves, however, there is observational reasoning for adopting the Gaussian function as the probability distribution of $\xi_{i}$; single-band quasar light curves are known to be well modelled by a first-order autoregressive Gaussian process known as a damped random walk \citep[e.g.,][and references therein]{mac12}.
This implies that $p(\xi_{i}|\psi)$ for magnitude$-$magnitude linear regression of quasar light curves can be modelled as a Gaussian distribution ($K = 1$), if the light curve is sampled for a sufficiently long duration of time [i.e., longer than the damping timescale of the time series, which is known to be $\sim$200 d in the rest frame for quasar variability, \citep[e.g.,][]{but11,and13}]. Therefore, in the following analyses, we assume $K = 1$, although the choice of the exact value of $K$ actually does not significantly affect the results obtained.
For $K=1$, $p(\xi_{i}|\psi)$ can be expressed as:
\begin{equation}
p(\xi_{i}|\psi) = \frac{1}{\sqrt{2\pi \tau^2}}\exp \left( -\frac{1}{2}\frac{(\xi_i - \mu)^2}{\tau^2} \right),
\label{informative_prior}
\end{equation}
where $\tau$ and $\mu$ are the parameters of this model.
This informative prior [equation~(\ref{informative_prior})] for sample positions enables Bayesian data analyses for linear regression of data with measurement errors on both axes \citep{gul89,kel07}.
The IDL routine for the Bayesian linear regression analysis discussed in \cite{kel07} ({\tt LINMIX\_ERR}) has been made available in the IDL Astronomy Users Library\footnote{http://idlastro.gsfc.nasa.gov/}.
The prior density distributions of the model parameters and the data likelihood function assumed in {\tt LINMIX\_ERR} can be found in \cite{kel07}.

The {\tt LINMIX\_ERR} estimator has an advantage over other non-Bayesian methods in that it calculates the posterior distribution of
the parameters for the given data using a Markov Chain Monte Carlo
(MCMC) method, and hence provides well-defined and reliable
parameter uncertainties \citep[see, e.g.,][]{par12}.
Therefore, we can estimate the intrinsic scatter in the magnitude$-$magnitude plot, denoted as $\sigma_{\text{int}}$(data) for the observation data (Section~\ref{data}) and as $\sigma_{\text{int}}$(model) for the model light curves (Section~\ref{simulation}), with their uncertainties, using {\tt LINMIX\_ERR}.

In addition to $\sigma_{\text{int}}$, we can also derive the single-band variability amplitude of the light curve plotted on the x-axis and its uncertainty by way of the model parameter $\tau$ appearing in equation~(\ref{informative_prior}).
%In addition to $\sigma_{\text{int}}$, as a model parameter $\tau$ appearing in equation~(\ref{informative_prior}), we can also derive the single-band variability amplitude of the light curve plotted on the $x$-axis and its uncertainty. 
By definition, $\tau$ is basically the same as the often-used variability amplitude indicator $V_{\text{std}}$, defined as $V_{\text{std}}=\sqrt{\Sigma^2 - \xi^2}$, where $\Sigma$ is the sample standard deviation and $\xi$ is the root mean square of the photometric error of a single-band light curve \citep[e.g.,][]{vau03,ses07,ai10,zuo12}.
In this work, we denote $\tau$ as $V(x)$:
\begin{equation}
V(x) = \tau,
\label{v_equal_tau}
\end{equation}
where $x$ indicates the photometric band corresponding to the $x$-axis
of a magnitude$-$magnitude plot (e.g., $x$ = $u$ for $u$-$g$, $u$-$r$, $u$-$i$ and $u$-$z$ plots), and refer to $V(x)$ as the variability amplitude. 
Fig.~\ref{fig:consistency} compares the variability amplitude of $g$-, $r$-, $i$- and $z$-band light curves of SDSS Stripe~82 quasars (see Section~\ref{data} for details) derived by the two
methods described above, i.e., $V_{\text{std}}$ and $V$. 
In Fig.~\ref{fig:consistency}, the values of $V$ are in good agreement with those of $V_{\text{std}}$.
This result validates the use of {\tt LINMIX\_ERR} as a reliable tool to quantify the quasar variability amplitude, and eventually its magnitude$-$magnitude correlation.

It should be noted that the broad-band photometric light curves
contain the flux variability not only of the accretion disc continuum
emission but also of the broad emission lines and the Balmer continuum emission \citep[e.g.,][]{bla82,obr95,bal95,kor01,pet04,kor04,wil05,cze13,kok14,che14,her14,ede15}.
However, contamination of emission line variability only makes the scatter in magnitude$-$magnitude space $\sigma_{\text{int}}$(data) larger because of the reverberation nature of the emission line variability \citep[e.g.,][]{sak11,sch12}.
In other words, we should consider $\sigma_{\text{int}}$(data) evaluated by the broad-band quasar light curves as an upper bound of the intrinsic scatter of the magnitude$-$magnitude correlation of the pure accretion disc continuum emission.
Therefore, even when the broad emission line variability dilutes the coherent variation of the pure accretion disc continuum emission to some extent, we can assess the invalidity of the Dexter \& Agol inhomogeneous accretion disc model by confirming the relation $\sigma_{\text{int}}$(model) $>$ $\sigma_{\text{int}}$(data).

\section{Data}
\label{data}

\begin{figure*}
\centering{
\includegraphics[clip, width=6.4in]{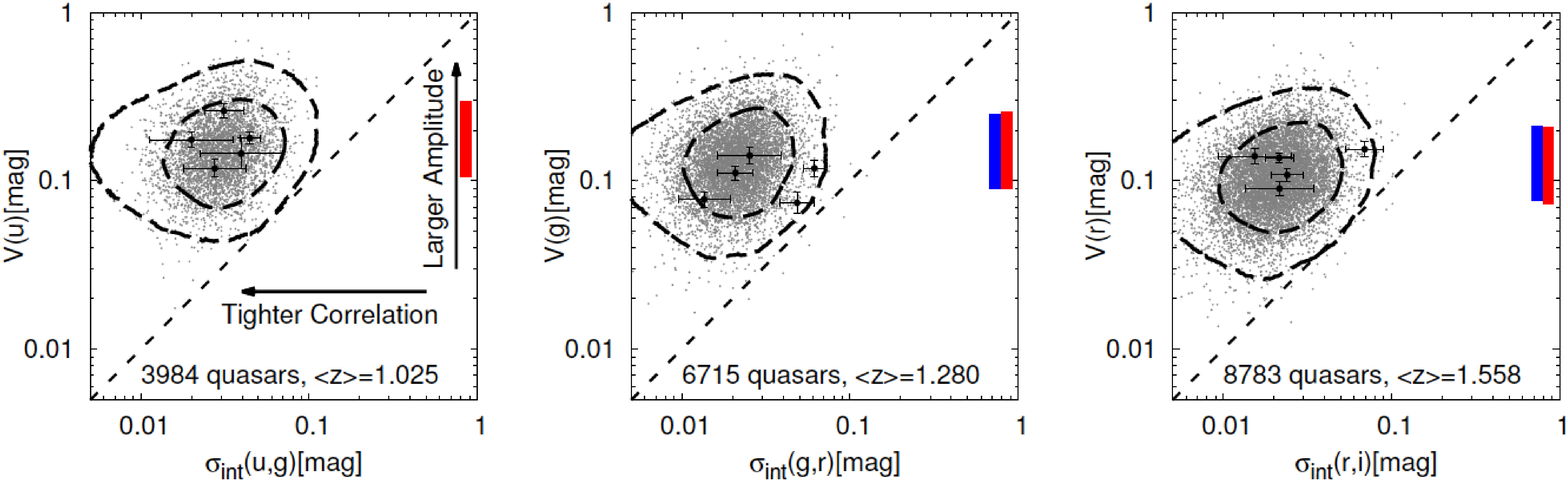}
}
\centering{
\includegraphics[clip, width=6.4in]{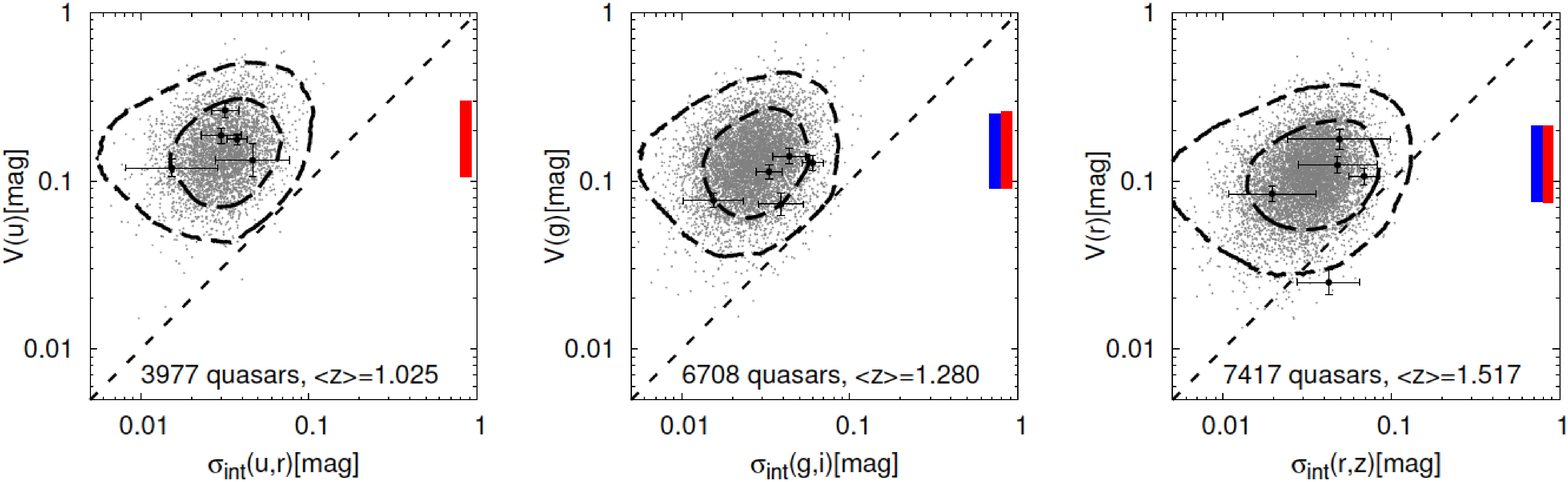}
}
\centering{
\includegraphics[clip, width=6.4in]{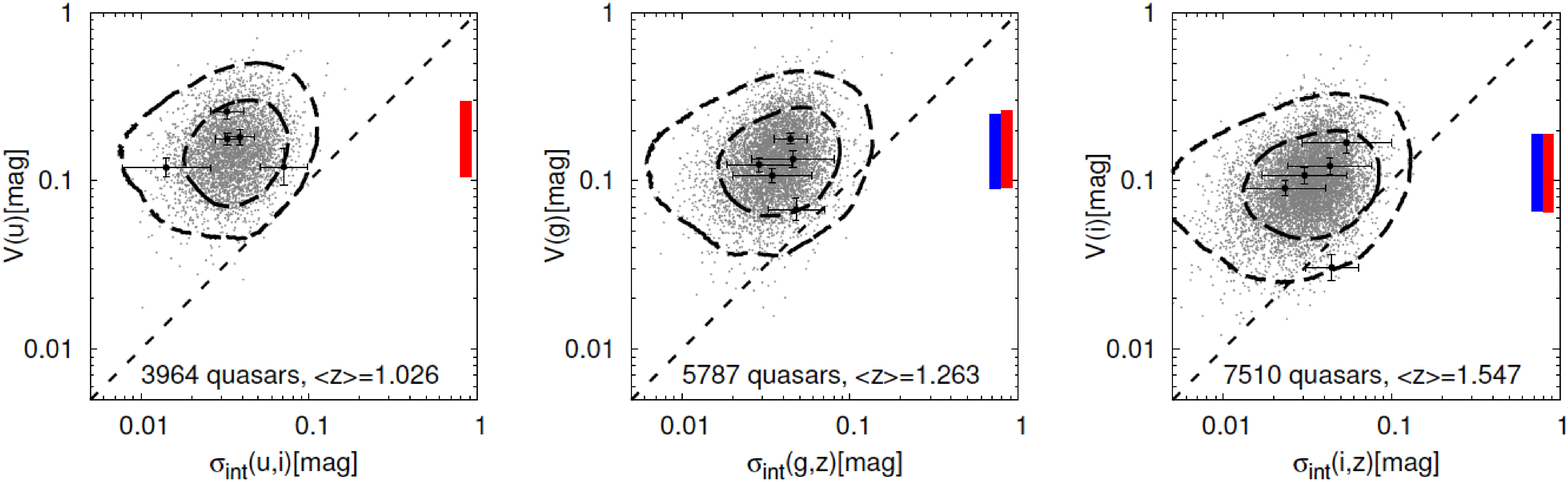}
}
\centering{
\includegraphics[clip, width=6.4in]{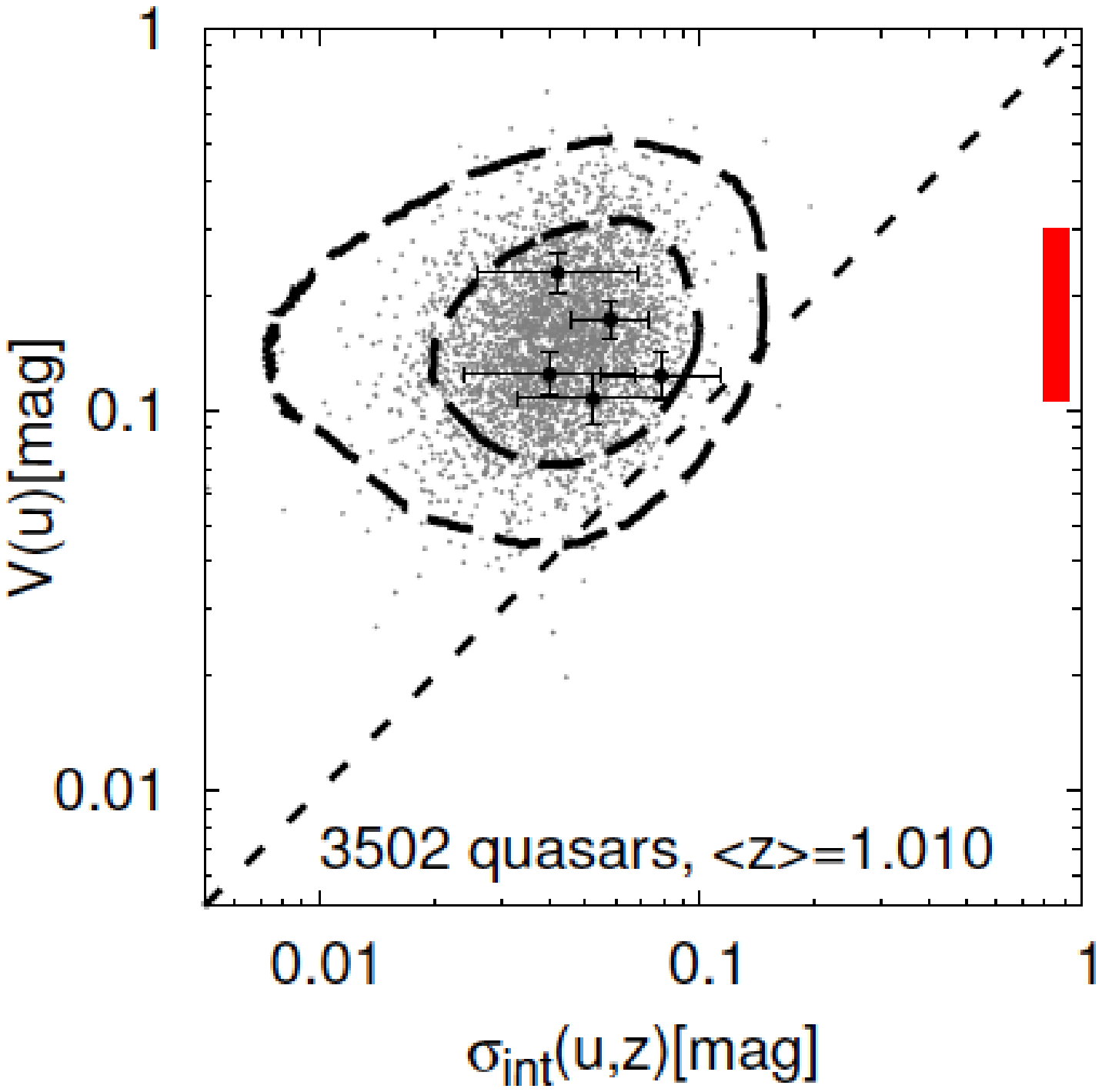}
}
\caption{Intrinsic scatter $\sigma_{\text{int}}$ versus variability amplitude $V$ for the SDSS Stripe~82 quasars obtained by the regression analyses in magnitude$-$magnitude space for all 10 band pairs (Section~\ref{data}).
Gray dots are the point estimates of the values of $\sigma_{\text{int}}$ and $V$.
Black points with error bars are the randomly selected points of $\sigma_{\text{int}}$ and $V$ to indicate the typical values of uncertainty in the point estimates. 
1$\sigma$ (68.2 per cent, inner dashed line) and 2$\sigma$ (95.4 per cent, outer dashed line) contours, which are calculated from a composite $\sigma_{\text{int}}$-$V$ posterior distribution constructed by summing each individual posterior distribution of the sample, are also shown in each of the panel. 
We show the range of median $\pm$ median absolute deviation of $V$ as red vertical bars on the right-hand side of the panels, and for comparison, we also show the median values of the variability amplitude (corresponding to $V_{\text{std}}$) derived by Zuo et al. 2012 for the $g$-, $r$- and $i$-band light curves of the SDSS Stripe~82 quasars as blue vertical bars on the left of the red vertical bars in the panels for $V(g)$, $V(r)$ and $V(i)$.
A thin dashed straight line in each of the panel indicates $V=\sigma_{\text{int}}$.
The number of objects and the mean redshift $\langle z \rangle$ for the sample of each band pair are indicated at the bottom of each panel.
}
\label{fig:data_locus}
\end{figure*}

Quantifying the intrinsic scatter in magnitude$-$magnitude space
requires simultaneous multi-epoch multi-band photometric light
curves for quasars. 
Hence, we use a database of the SDSS Stripe~82 multi-epoch five-band light curves for spectroscopically confirmed quasars from the fifth SDSS quasar catalogue \citep{sch10} presented by \cite{mac12}.
SDSS Stripe 82 region is located in the centre of the three stripes in the South Galactic Cap and has been imaged with $u$, $g$, $r$, $i$ and $z$ SDSS filters simultaneously about 60 times on average from 1998 to 2007 \citep[][]{aba09}.
The average wavelengths for the $u$-, $g$-, $r$-, $i$- and $z$-bands
are 3551, 4686, 6165, 7481 and 8931\AA, respectively\footnote{http://www.sdss.org/dr7/}.
We cross-match the Stripe~82 light curve catalogue of \cite{mac12} with a catalogue of quasar properties from SDSS DR7 \citep{she11}, and use the improved red shift estimates \citep{hew10} listed in \cite{she11}'s catalogue as the redshift of each quasar.

For completeness, we use all the five-band light curves.
The $x$-axis of the two-band magnitude$-$magnitude plot ($x$-$y$ space) is always chosen to be the shorter wavelength band of the
two. 
This results in a total of 10 band pairs: $u$-$g$, $u$-$r$, $u$-$i$,
$u$-$z$, $g$-$r$, $g$-$i$, $g$-$z$, $r$-$i$, $r$-$z$ and $i$-$z$ magnitude$-$magnitude plots. 
To avoid the Lyman $\alpha$ absorption ($<$1216 \AA) entering into the $u$-, $g$-, $r$- and $i$-bands, we focus only on the redshift range below $z$ $=$ 1.5, 2.0, 3.5 and 4.5 for the regression analyses of
band pairs containing the $u$-, $g$-, $r$- and $i$-bands, respectively
(as listed in Table~\ref{tbl:data_properties}).

It is known that the SDSS Stripe~82 light curves contain some outlying photometric points that are several magnitudes fainter (or brighter) than adjacent values, most of which must have a non-physical origin \citep[e.g.,][]{schmidt10,mac10,meu11,pal11,zuo12}.
To eliminate these outlying measurements, a medianized light curve was generated by applying a seven-point median filter to each of the light curves of the five bands, and then the measurements with a residual between the medianized light curve and the photometric data larger than 0.25 mag were removed \citep[following the procedures of ][]{schmidt10}.
Then, we exclude quasar light curves that show no variability, i.e., whose sample variance is smaller than the mean measurement error variance \citep[e.g.,][]{zuo12}.
In addition, since we are attempting to focus on the properties of long-term and multi-epoch variability, we exclude from our sample quasar light curves that have less than 20 photometric epochs.
The mean total photometric points $\langle N_{\text{epoch}} \rangle$ in the resulting two-band light curves are listed in Table~\ref{tbl:data_properties}. 
Then, linear regression analyses in magnitude$-$magnitude space for these two-band light curves are conducted using {\tt LINMIX\_ERR}.
The potential scale reduction factor $\hat{R}$ \citep{gel92} is used in {\tt LINMIX\_ERR} to monitor the convergence of the MCMC to the posterior, and we do not use data for which the condition $\hat{R}$ $<$ $1.1$ is not satisfied within 10 000 iterations of MCMC. 
The final quasar sample size $N_{\text{obj}}$ is listed in Table~\ref{tbl:data_properties}.

In Fig.~\ref{fig:data_locus}, we plot the results of the regression analyses for the SDSS Stripe~82 data in magnitude$-$magnitude space; i.e., the intrinsic scatter $\sigma_{\text{int}}$ versus variability amplitude $V$ for all 10 band pairs.
Point estimates of the values of $\sigma_{\text{int}}$ and $V$ for each individual quasar are defined as the mean and the standard deviation of the posterior distributions of these parameters. 
1$\sigma$ (68.2 per cent) and 2$\sigma$ (95.4 per cent) contours of the $\sigma_{\text{int}}$-$V$ relation for the SDSS Stripe~82 data, which we use as the observational constraint in later sections, are calculated from a composite $\sigma_{\text{int}}$-$V$ posterior distribution constructed by summing each individual posterior distribution of the sample quasars.

A red vertical bar on the right-hand side in each of the panels of
Fig.~\ref{fig:data_locus} shows the range of the median $\pm$ median absolute deviation of $V$. 
And for comparison, we also show the median values of the variability amplitude (corresponding to $V_{\text{std}}$) derived by \cite{zuo12} for the $g$-, $r$- and $i$-band light curves of the SDSS Stripe~82 quasars as blue vertical bars on the left of the red vertical bars in the panels for $V(g)$, $V(r)$ and $V(i)$.
In Fig.~\ref{fig:data_locus}, we can see that the values of the variability amplitude $V_{\text{std}}$ derived by \cite{zuo12} are consistent with those of $V$ derived by {\tt LINMIX\_ERR}.
Again (as already mentioned in Section~\ref{method}), this consistency validates the use of {\tt LINMIX\_ERR} as a reliable tool for quantifying the quasar magnitude$-$magnitude correlation.

As indicated in Fig.~\ref{fig:data_locus}, smaller values of $\sigma_{\text{int}}$ indicate tighter linear correlation in magnitude$-$magnitude space, and larger values of $V$ indicate a larger variability amplitude.
The 1$\sigma$ and 2$\sigma$ contours generally satisfy $V$ $\gg$ $\sigma_{\text{int}}$, which means that the two-band light curves of quasars are highly correlated.

\section{Inhomogeneous Accretion Disc Model}
\label{IADmodel}

\begin{table*}
\caption{Assumed photometric errors in the model five-band light curves (in magnitude unit).}
\label{tbl:model_error}
\begin{tabular}{@{}cccccccccc}
\hline
Group & 1 & 2 & 3 & 4 & 5 & 6 & 7 & 8 & 9 \\
\hline
$\sigma_u$&   $0.018$&$0.021$&$0.028$&$0.037$&$0.051$&$0.071$&$0.097$&$0.14$&$0.20$ \\ 
$\sigma_g$&   $0.011$&$0.012$&$0.014$&$0.016$&$0.021$&$0.027$&$0.035$&$0.048$&$0.068$ \\ 
$\sigma_r$&   $0.0099$&$0.011$&$0.013$&$0.015$&$0.019$&$0.025$&$0.034$&$0.047$&$0.066$ \\ 
$\sigma_i$&   $0.011$&$0.012$&$0.014$&$0.017$&$0.022$&$0.029$&$0.040$&$0.057$&$0.081$ \\ 
$\sigma_z$&   $0.016$&$0.020$&$0.027$&$0.037$&$0.054$&$0.079$&$0.11$&$0.16$&$0.25$ \\
Per cent& $1$&$2$&$4$&$10$&$18$&$28$&$28$&$8$&$1$ \\ 
\hline
\end{tabular}
\end{table*} 

\begin{figure}
\center{
\includegraphics[clip, width=2.95in]{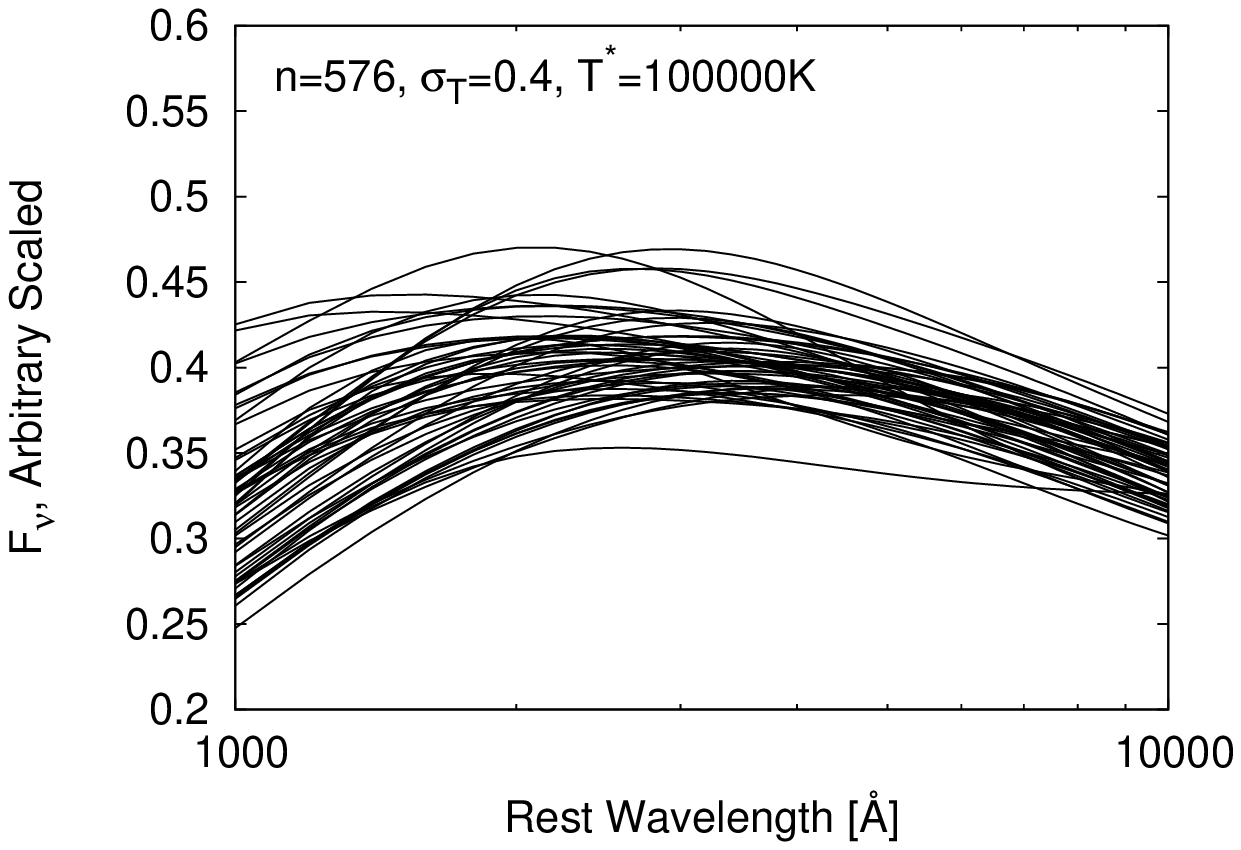}
}
\center{
\includegraphics[clip, width=2.95in]{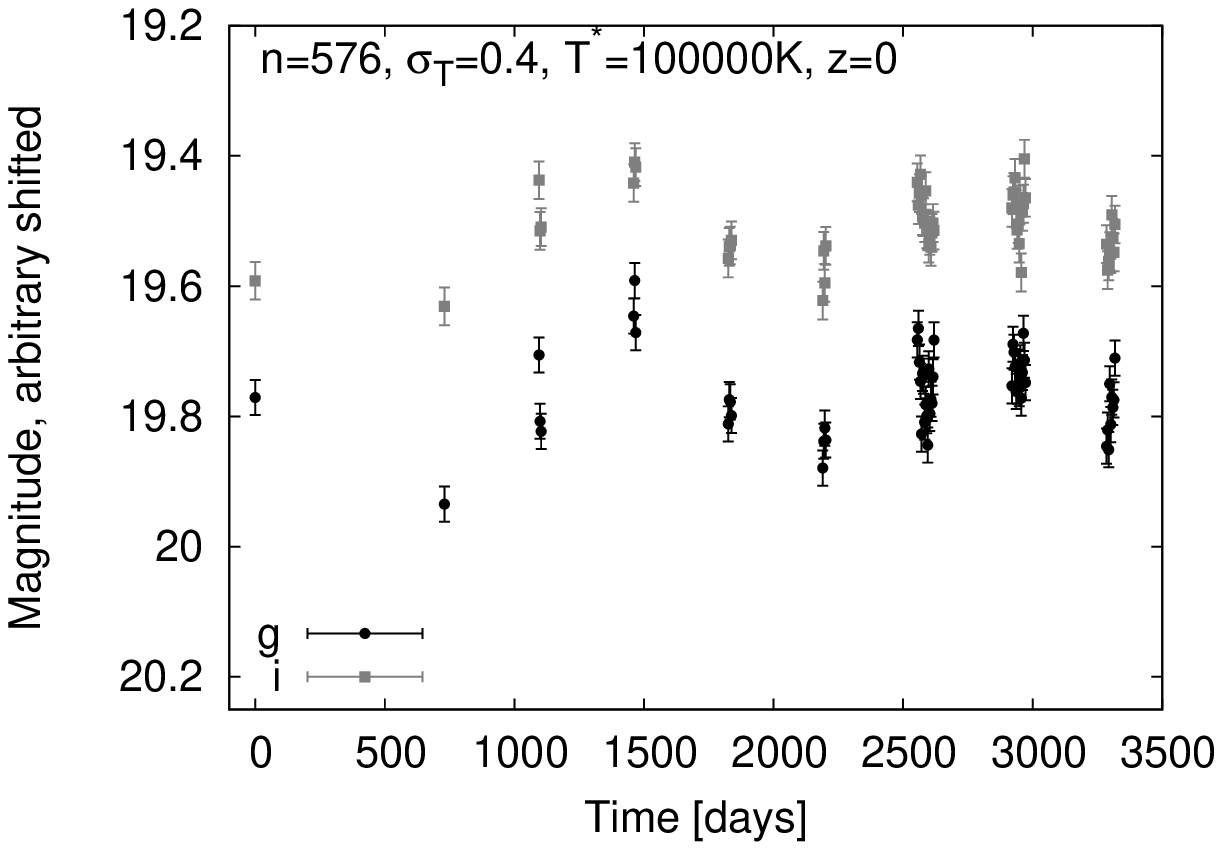}
}
\center{
\includegraphics[clip, width=2.75in]{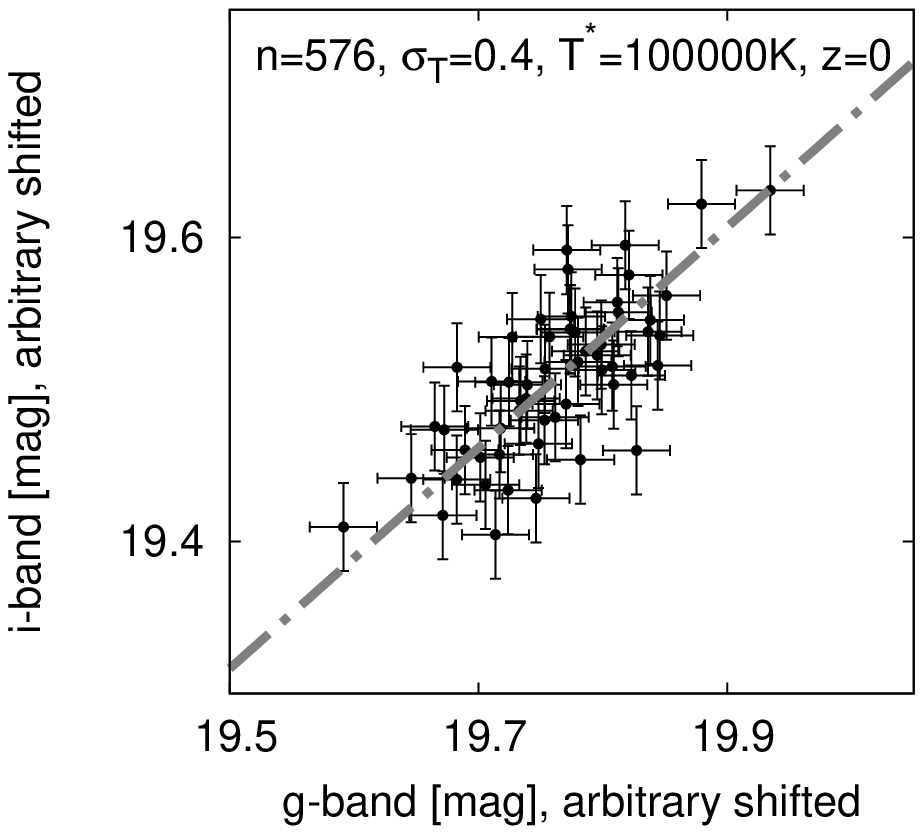}
}
\caption{
Top: A single realization of the 10-yr model spectral time series, assuming $\sigma_{T}$ = 0.40, $n$ = 576 and $T^*=100 000$ K. 
Middle: $g$- and $i$- band model light curves obtained from the spectral time series shown in the top panel, assuming $z = 0$, $\sigma_{g}=0.027$ mag and $\sigma_{i}=0.029$ mag.
Bottom: Model magnitude$-$magnitude plot of the $g$-$i$ band pair light curve shown in the middle panel. The straight line indicates the linear regression line drawn using {\tt LINMIX\_ERR}.
}
\label{fig:2}
\end{figure}

\subsection{Dexter \& Agol's formalization}
\label{formalization}

Steady-state black hole accretion discs residing in luminous AGNs or quasars are thought to be well described by the standard thin accretion disc model \citep{sha73,nov73}.
The emitted spectrum from the standard thin accretion disc can be written as a sum of blackbody spectra from the surface elements of the disc \citep[e.g.,][]{fra92,kat08}:
\begin{eqnarray}
F_{\nu}(\nu_{\text{obs}})&\propto & \nu^3\int_{}^{}\frac{1}{e^{h\nu /k_B T_{\text{eff}}(R)}-1}dS\\
T_{\text{eff}}(R) &=& \left( \frac{3GM_{\text{BH}}\dot{M}}{8\pi \sigma R_{\text{in}}^3} \right)^{1/4} \left(\frac{R_{\text{in}}}{R}\right)^{3/4} \left(1-\sqrt{\frac{R_{\text{in}}}{R}}\right)^{1/4} \nonumber \\
&= & T^* \left(\frac{1}{x}\right)^{3/4} \left(1-\sqrt{\frac{1}{x}}\right)^{1/4},
\label{standard_temperature}
\end{eqnarray}
where $h$, $k_{B}$, $G$, and $\sigma$ are Planck constant, Boltzmann constant, gravitational constant and the Stefan-Boltzmann constant, respectively.
$\nu_{\text{obs}}$ is the observed-frame frequency, which is related to the rest-frame frequency $\nu$ as $\nu_{\text{obs}} = \nu/(1+z)$.
$M_{\text{BH}}$ and $\dot{M}$ represent the black hole mass and the mass accretion rate, respectively.
$dS = Rd\phi dR$ is the surface element of the accretion disc, where $R$ and $\phi$ are the radial and azimuthal coordinates on the disc plane, respectively.
The disc inner radius $R_{\text{in}}$ is taken to be $R_{\text{in}}=3R_{S}$ where $R_{S}$ is the Schwarzschild radius $R_{S}=2GM_{\text{BH}}/c^2$.
$3R_{S}$ corresponds to the innermost stable circular orbit of a non-spinning Schwarzschild black hole.
The dimensionless radial coordinate $x$ is defined as $x=R/R_\text{in}$, which means the disc inner radius is $x_{\text{in}}=1$.
\begin{equation}
T^* = \left(\frac{3GM_{\text{BH}}\dot{M}}{8\pi \sigma R_{\text{in}}^3}\right)^{1/4}
\end{equation}
is the disc characteristic temperature, which is the only parameter determining the spectral shape of the emitted spectrum $L_{\nu}$ from the standard thin accretion disc \citep[][]{per06,kok14,rua14}.
For the SDSS quasars, $T^{*}$ ranges from $50 000$ to $200 000$ K if the Newtonian value of radiative efficiency ($\epsilon = 1/12$) is adopted \citep{kok14}.

It should be noted that there are several pieces of evidence from the studies of hot-dust-poor AGNs, dust reverberation mapping and infrared polarimetry, suggesting that emission from quasar accretion discs extends well into near-infrared wavelengths, which means that an accretion disc does not truncate even at a few thousands of $R_S$ \citep[e.g.,][and references therein]{kis05,tom06,kis07,kis08,hon10,hao10,lir11,kos14,okn14}.
Therefore, we fix the disc outer radius as $x_{\text{out}}=2^{16}$ for convenience of the model calculations. 
Since emissions from the outer region of the quasar accretion disc (e.g., $x>10 000$) are negligible in the UV-optical wavelength range, the exact choice of the disc outer radius has little effect on the resulting UV-optical spectra.

\cite{dex11} and \cite{rua14} calculated inhomogeneous accretion disc model spectra by dividing the disc into $n$ zones per factor of 2 in radius.
The zones are log-spaced in $r$ and evenly spaced in $\phi$.
From an $N$ $\times$ $N$ grid with inner and outer radii of $x_{\text{in}}=1$ and $x_{\text{out}}=2^{16}$, the parameter $n$ is given by 
\begin{equation}
n=N\frac{\text{log}(2)}{\text{log}(2^{16})}\times N = \frac{N^2}{16}. \nonumber
\end{equation}
To cover the whole range of the parameter space suggested by \cite{dex11} sufficiently (i.e., $n$ $=$ 10$-$1 000), in this work
we consider the cases $n$ $=$ 16, 64, 144, 256, 400, 576, 784, 1024,
1296, 1600, 1936, 2304, 2704, 3136, 3600 and 4096.

In Dexter \& Agol's inhomogeneous accretion disc model, disc inhomogeneity is given by adding temperature fluctuations to the effective temperature profile [equation~(\ref{standard_temperature})] for each zone independently.
The effective temperature of each zone $T_{\text{eff}}$ ($x$, $\phi$) is assumed to undergo a damped random walk \citep[e.g.,][]{kel09,koz10,mac10,her14} with an amplitude $\sigma_T$ (in the unit of  $\log T_{\text{eff}}$); 
in the limit $n\rightarrow \infty$, the damped random walk model becomes time independent with a log-normal distribution of disc temperatures in each annulus whose variance is ($\ln 10$ $\sigma_T)^2/2$  [equation~(2) of \citep{dex11}]. 
As assumed in \cite{dex11} and \cite{rua14}, the characteristic decay timescale of the temperature fluctuations $\tau_{\text{damp}}$ is fixed to be 200 d, although the results are insensitive to the choice of $\tau_{\text{damp}}$ \citep{dex11}.

\subsection{Simulating the SDSS Stripe 82 quasar light curves}
\label{simulation}

\begin{figure}
\center{
\includegraphics[clip, width=3.2in]{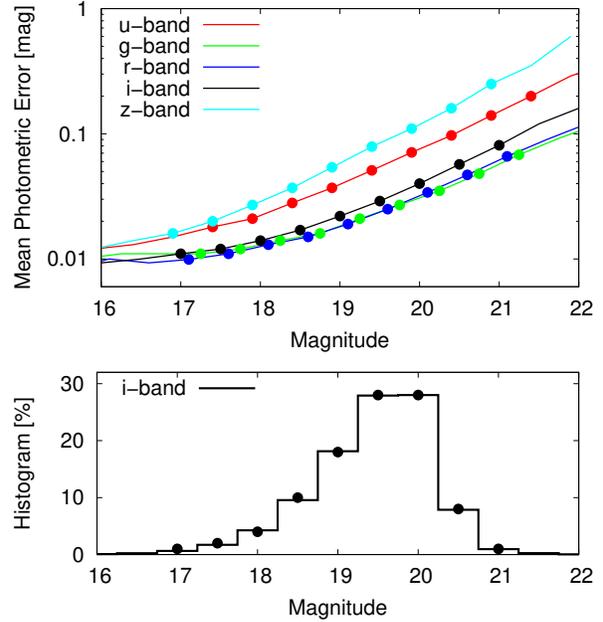}
}
\caption{
Mean photometric errors of the SDSS Stripe~82 quasar light curves as a function of the magnitude (upper panel), and a histogram of the $i$-band magnitudes of the quasar light curves (lower panel).
The photometric errors and the percentages of the assignment into nine groups assumed in the model calculations (listed in Table~\ref{tbl:model_error}) are shown as points in the upper and the lower panels, respectively (see Section~\ref{simulation} for details).
}
\label{fig:photometric_error}
\end{figure}

The Dexter \& Agol inhomogeneous accretion disc model is characterized by two model parameters, $n$ and $\sigma_{T}$, which represent the number of separated zones in an annulus of the disc and the amplitude of temperature fluctuations, respectively. 
In addition to these two parameters, the spectral shape depends on a physical parameter, $T^{*}$ \citep{per06,kok14,rua14}.
The quasar redshift $z$ determines which of the rest-frame wavelengths fall on to the broad-band filters. 
In summary, there are four model parameters in our model calculations: $n$, $\sigma_{T}$, $T^*$, and redshift $z$.
\cite{dex11} concluded that, for quasar accretion discs, the two parameters $n$ and $\sigma_{T}$ should be in the range of $n$ $=$ 100$-$1 000 and $\sigma_{T}$ $=$ 0.35$-$0.50 to fit several observational constraints, including microlensing accretion disc size measurements, quasar UV spectral shape and the UV-optical variability amplitude (see Section~\ref{intro}).
\cite{rua14} showed that the composite difference spectrum of the SDSS quasars can be fitted by the Dexter \& Agol inhomogeneous accretion disc model with the same parameter ranges as \cite{dex11}.
We decide to explore the parameter space of 16 $\leq$ $n$ $\leq$ 4096 and 0.1 $\leq$ $\sigma_{T}$ $\leq$ 0.6, which sufficiently covers the
whole range of the parameter space suggested by \cite{dex11}.

The parameters $T^{*}$ and $z$ are taken to represent the SDSS Stripe~82 data. 
For the SDSS quasars, the parameter $T^{*}$ ranges from 50 000 K
to 200 000 K \citep{kok14,rua14}. 
The redshift $z$ is taken to be $z$ $<$ 1.5 for $u$-$g$, $u$-$r$, $u$-$i$ and $u$-$z$ band pairs; $z$ $<$ 2.0 for $g$-$r$, $g$-$i$ and $g$-$z$ band pairs; $z$ $<$ 3.5 for $r$-$i$ and $r$-$z$ band pairs;
and $z$ $<$ 4.5 for the $i$-$z$ band pair (see Section~\ref{data}).

For a fair comparison between the Dexter \& Agol inhomogeneous disc model and the SDSS Stripe~82 quasar light curves, we have to calculate mock light curves that adequately model the actual SDSS Stripe~82 data.
The number of measurement points and their sampling intervals assumed in the model light curves are determined to make them similar to the average SDSS Stripe~82 light curve;
i.e., we assume that we have 10 yr of light curves with 1, 0, 1, 3, 3, 4, 4, 17, 14 and 9 measurement points in each of the 10 yr (56 points in total), and the sampling interval within each of the years is fixed to 4 d (see the middle panel of Fig.~\ref{fig:2}) \citep[e.g.,][]{sak14}.
In the actual SDSS Stripe~82 data, the first 7 yr are the runs obtained as part of the SDSS-I Legacy survey \citep{ade07}, and the last 3 yr correspond to the SDSS-II Supernovae project runs \citep{fri08}.

Moreover, for a fair comparison between the model and the data, Gaussian photometric errors are added to the model light curves and the same linear regression method (i.e., {\tt LINMIX\_ERR}) is applied to the photometric error-added model magnitude$-$magnitude plots. 
We calculate the 100 realizations of the model five-band light curves for each of the parameter sets ($n$, $\sigma_T$, $T^*$, $z$), and then divide them into nine groups and assign the Gaussian photometric errors to each of the nine groups as tabulated in Table~\ref{tbl:model_error}.
The photometric errors and the percentages of the assignment into nine groups are determined to make them comparable to the histogram of the photometric errors of the SDSS Stripe 82 quasar light curves. 
Fig.~\ref{fig:photometric_error} shows the mean photometric errors of the SDSS Stripe~82 quasar light curves as a function of the magnitude. First, nine representative values of the photometric errors for the $i$-band (black points in the upper panel of Fig.~\ref{fig:photometric_error}) are sampled according to the histogram of the $i$-band magnitudes of the quasar light curves rounded to a whole number (black points in the lower panel of Fig.~\ref{fig:photometric_error}).
The percentages of the assignment into nine groups are fixed to those of the $i$-band values (black points in the lower panel of Fig.~\ref{fig:photometric_error}), as listed in the last row
of Table~\ref{tbl:model_error}. 
Then, the $i$-band magnitude histogram is shifted in the magnitude direction assuming the mean quasar spectral index of $f_{\nu} \propto \nu^{-0.5}$ \citep{van01}, and the photometric errors for the $u$-, $g$-, $r$- and $z$-bands are sampled according to the shifted histograms (upper panel of Fig.~\ref{fig:photometric_error}).
The values of the photometric errors assumed in the model light curves for each of the nine groups are listed in Table~\ref{tbl:model_error}. Each of the light curves is assumed to have fixed values of photometric errors; i.e., we ignore the changes in photometric errors within each of the light curves. 
Finally, we apply the same exclusion criteria for the simulated light curves as those for the SDSS Stripe~82 data described in Section~\ref{data}.

It should be noted that the assumed values of the sampling numbers and the sampling intervals described above do not have significant impacts on the regression results, since the detailed information of the sampling cadence is essentially abandoned in the linear regression analyses in magnitude$-$magnitude space. 
Moreover, since {\tt LINMIX\_ERR} includes the proper models of the Gaussian photometric errors, the exact values of the assumed photometric errors also do not significantly affect the regression results.

The top panel of Fig.~\ref{fig:2} shows a single realization of the 10 yr model spectral time series, assuming $\sigma_{T}$ = 0.40, $n$ = 576 and $T^{*}$ = 100 000 K. 
As expected, the superposition of independent temperature fluctuations on the accretion disc makes the spectral shape of the model spectra highly variable. 
The middle panel of Fig.~\ref{fig:2} shows the $g$- and $i$- band model light curves obtained from the spectral time series shown in the top panel of Fig.~\ref{fig:2}, assuming $z = 0$, $\sigma_{g}=0.027$ mag and $\sigma_{i}=0.029$ mag, where $\sigma_{g}$ and $\sigma_{i}$ are the assumed photometric errors.
The bottom panel of Fig.~\ref{fig:2} shows the resulting regression line for the $g$-$i$ band pair magnitude$-$magnitude plot of the two-band light curves of the middle panel. 
A quantitative discussion on the linear regression results for various
values of model parameters are given in the next section.

\section{Comparison of the Inhomogeneous Accretion Disc Model with Observational Data}
\label{model_comparison}

\begin{figure}
\center{
\includegraphics[clip, width=5.5in, angle=90]{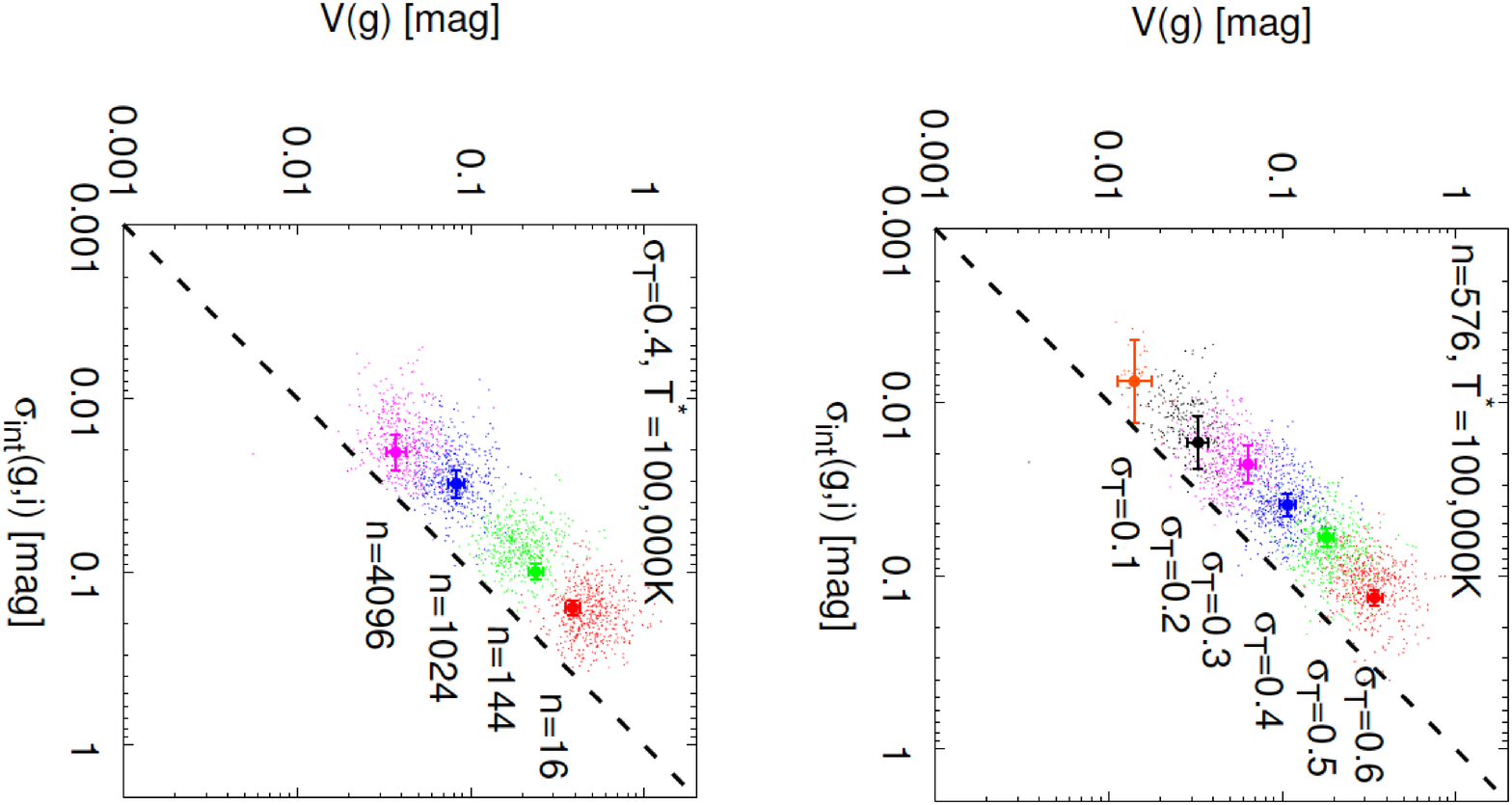}
}
\caption{
Intrinsic scatter $\sigma_{\text{int}}$ versus variability amplitude $V$ for the Dexter \& Agol inhomogeneous accretion disc model for $g$-$i$ band pairs (dots in different colours). 
Coloured points with error bars are the randomly selected points of $\sigma_{\text{int}}$ and $V$ to indicate the typical values of uncertainty in the point estimates. 
Top: $\sigma_{\text{int}}$-$V$ relation for various values 0.1 $\leq$ $\sigma_T$ $\leq$ 0.6 (in 0.1 increments; in different colours). 
All redshifts, $z$ $=$ 0.0, 0.5, 1.0, 1.5 and 2.0, are plotted simultaneously, and the other model parameters are fixed to $n$ $=$ 576 and $T^{*}$ $=$ 100 000 K. 
There are fewer points for smaller values of $\sigma_{T}$ because we have excluded the model light curves whose flux variability is smaller compared to the assumed photometric errors (see Sections~\ref{data} and \ref{simulation} for details). 
Bottom: The same as the top panel, but for various values $n$ $=$ 16, 144, 1024 and 4096 (in different colours).
The other model parameters are fixed to $\sigma_{T}$ $=$ 0.4 and $T^{*}$ $=$ 100 000 K. 
A thin straight line in each of the panels indicates $\sigma_{\text{int}} = V$.
}
\label{fig:large_sigma}
\end{figure}

Fig.~\ref{fig:large_sigma} shows the intrinsic scatter $\sigma_{\text{int}}$ versus variability amplitude $V$ for the Dexter \& Agol inhomogeneous accretion disc model for the $g$-$i$ band pair, for various values of $\sigma_{T}$ in 0.1 increments (upper panel), and for various values of $n$ $=$ 16, 144, 1024 and 4096 (bottom panel). 
Since large temperature fluctuations lead to large flux variability, the variability amplitude $V$ becomes larger when $\sigma_{T}$ becomes larger. 
On the other hand, since the sum of a large number of independent fluctuations lessens the stochastic behaviour, the variability amplitude becomes smaller when $n$ becomes larger [see equation~(1) of \cite{dex11}]. 
In the upper panel of Fig.~\ref{fig:large_sigma}, it is apparent that there are fewer points for smaller values of $\sigma_{T}$; this is because we have excluded the model light curves whose flux variability is smaller compared to the assumed photometric errors (see Sections~\ref{data} and \ref{simulation}).

In Fig.~\ref{fig:large_sigma}, the intrinsic scatter $\sigma_{\text{int}}$ is only slightly smaller than the variability amplitude $V$ (i.e., $\sigma_{\text{int}}$ $\lesssim$ $V$), which means that the Dexter \& Agol inhomogeneous accretion disc model predicts a weak magnitude$-$magnitude correlation. 
Moreover, as we can see in Fig.~\ref{fig:large_sigma}, the $\sigma_{\text{int}}$-$V$ relation for various values of $\sigma_{T}$ and $n$ follows a linear track from left bottom to top right, which indicates that the weakness of the magnitude$-$magnitude correlation is kept for the whole of the model parameter space considered here. 
Since (as shown in Fig.~\ref{fig:large_sigma}) the model values of $\sigma_{\text{int}}$ and $V$ for various model parameters are confined within a certain locus in $\sigma_{\text{int}}$-$V$ space, and we are not interested in the details of the model parameter dependence of them, hereafter we refer to the $\sigma_{\text{int}}$-$V$ relation of the Dexter \& Agol inhomogeneous accretion disc model as the region in the $\sigma_{\text{int}}$-$V$ space defined by all the model values of $\sigma_{\text{int}}$ and $V$ for all the values of the model parameters, as described below.

\begin{figure*}
\leftline{
\includegraphics[clip, width=2.3in]{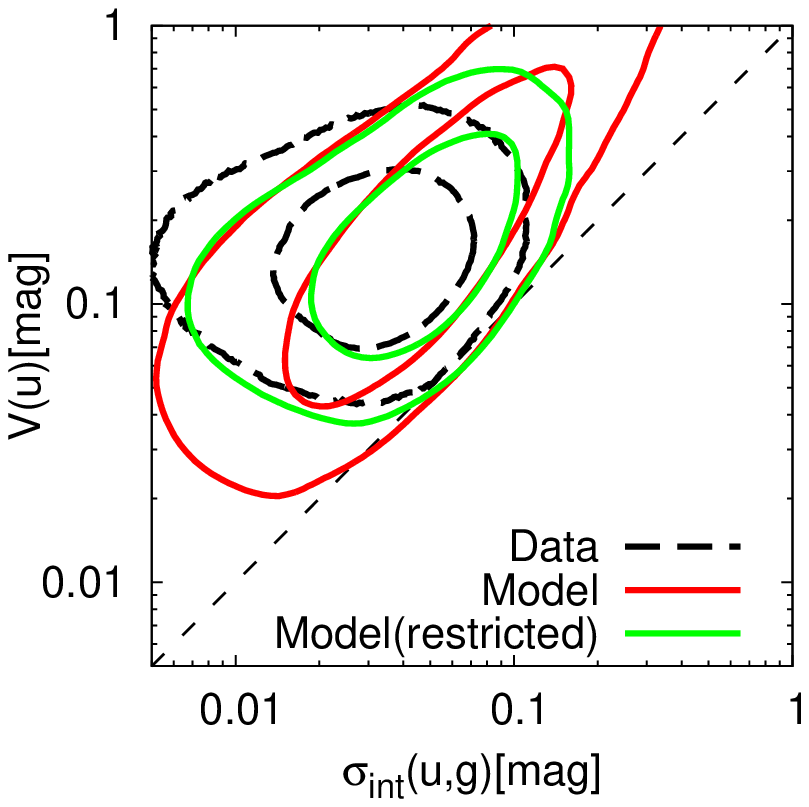}
\hspace{-0.4cm}
\vspace{-0.1cm}
\includegraphics[clip, width=2.3in]{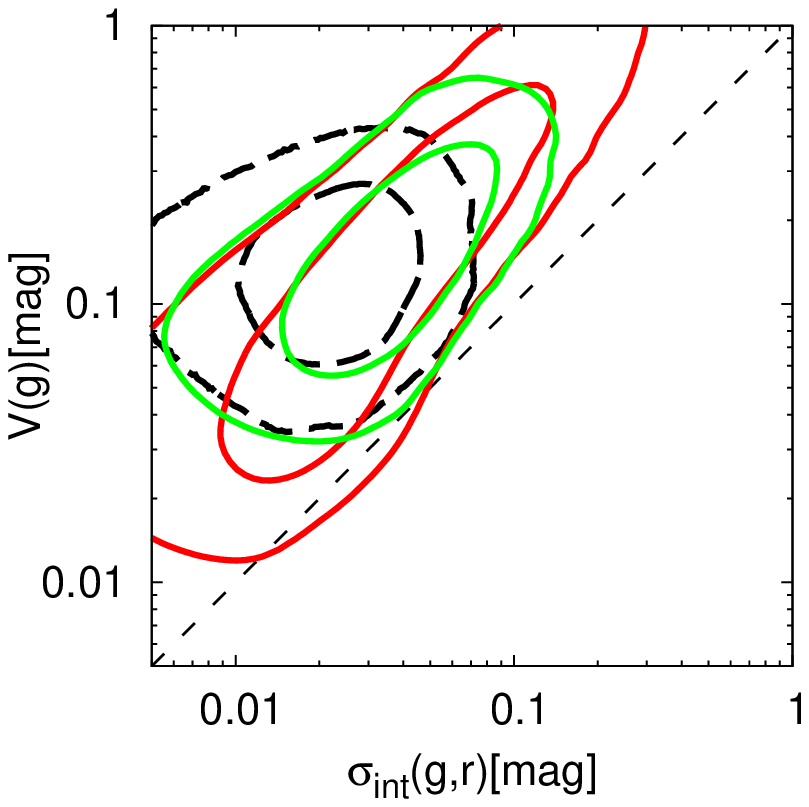}
\hspace{-0.4cm}
\includegraphics[clip, width=2.3in]{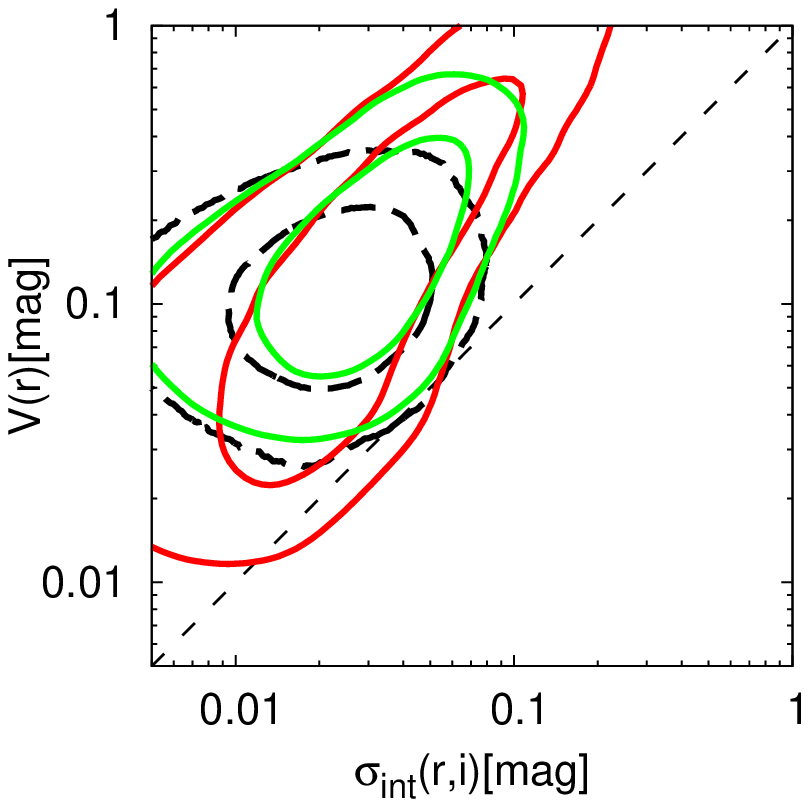}
\hspace{-0.4cm}
}
\leftline{
\includegraphics[clip, width=2.3in]{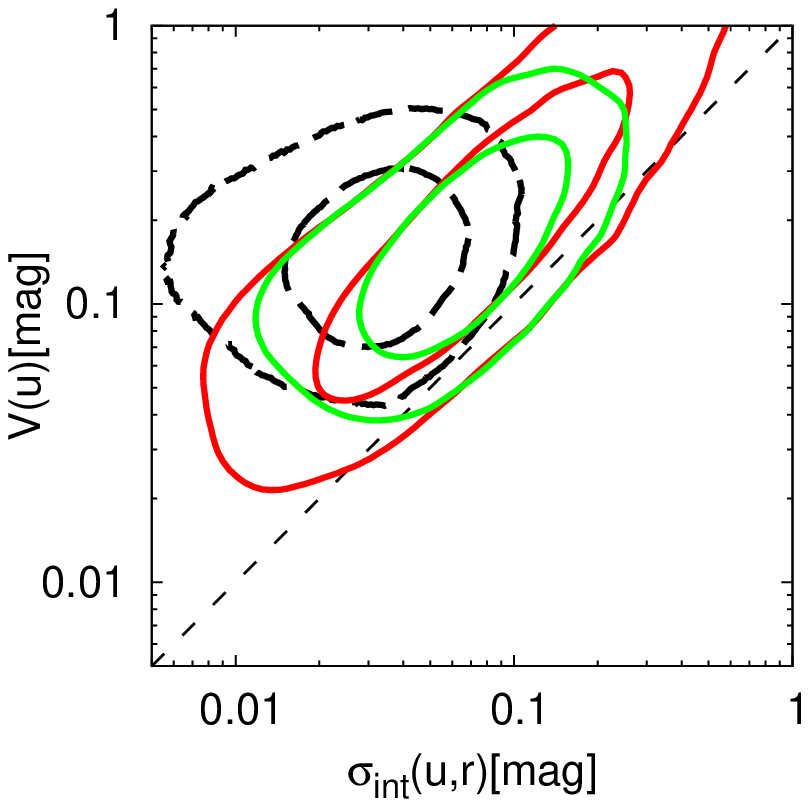}
\hspace{-0.4cm}
\vspace{-0.1cm}
\includegraphics[clip, width=2.3in]{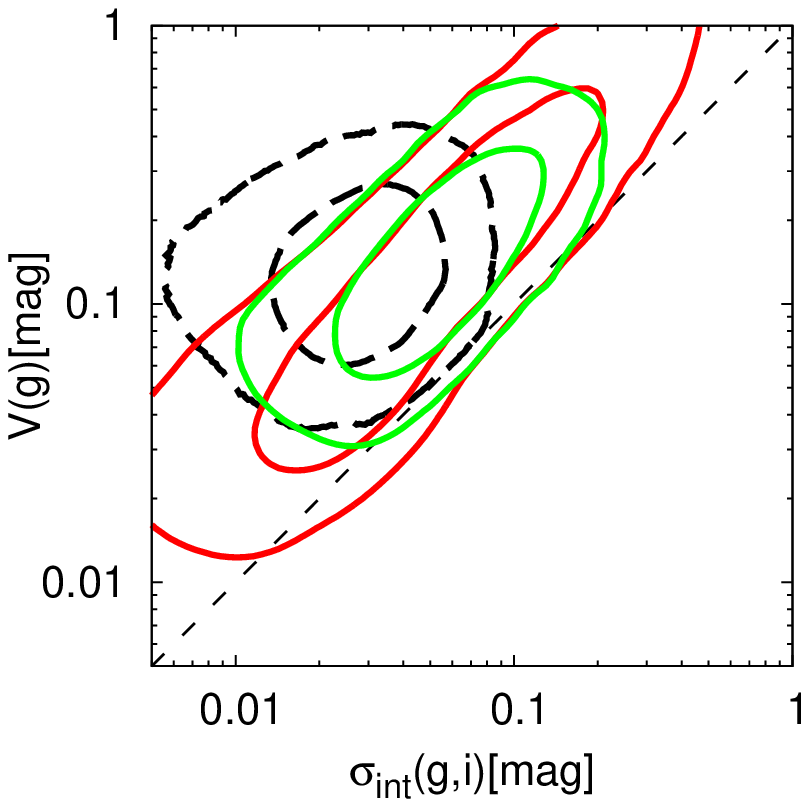}
\hspace{-0.4cm}
\includegraphics[clip, width=2.3in]{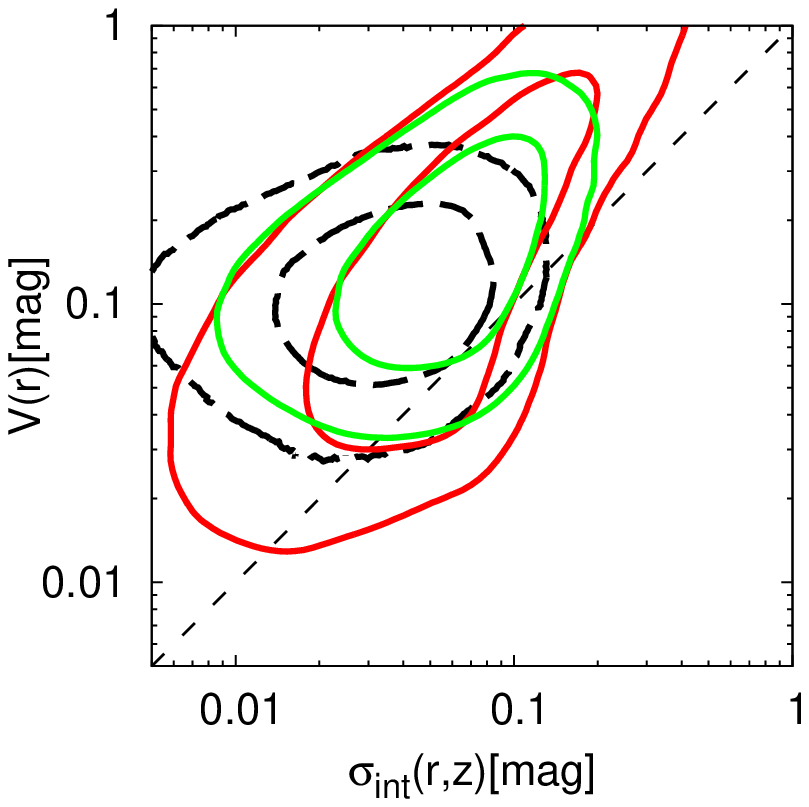}
\hspace{-0.4cm}
}
\leftline{
\includegraphics[clip, width=2.3in]{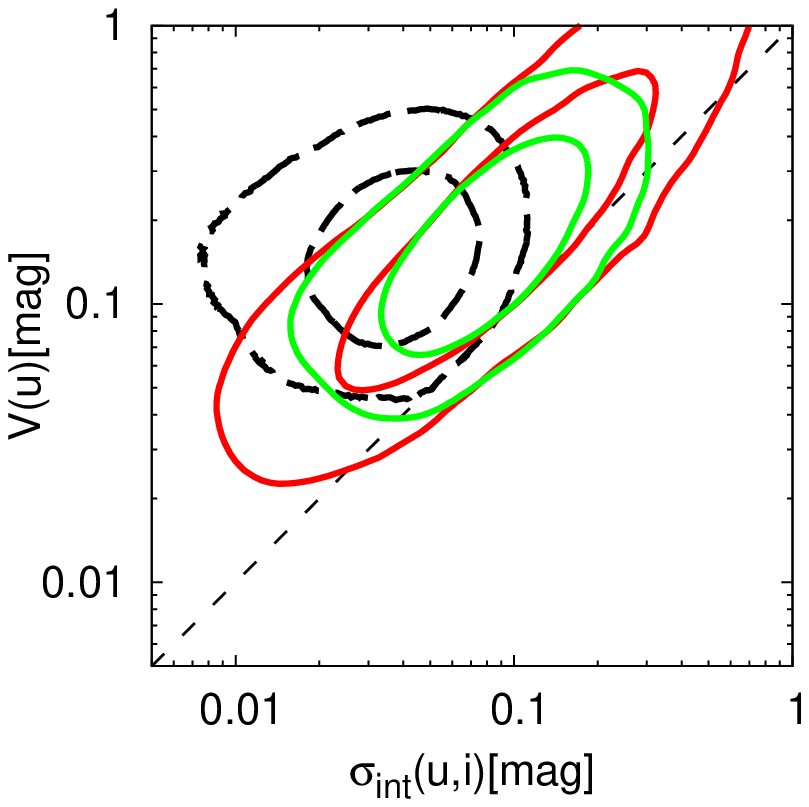}
\hspace{-0.4cm}
\vspace{-0.1cm}
\includegraphics[clip, width=2.3in]{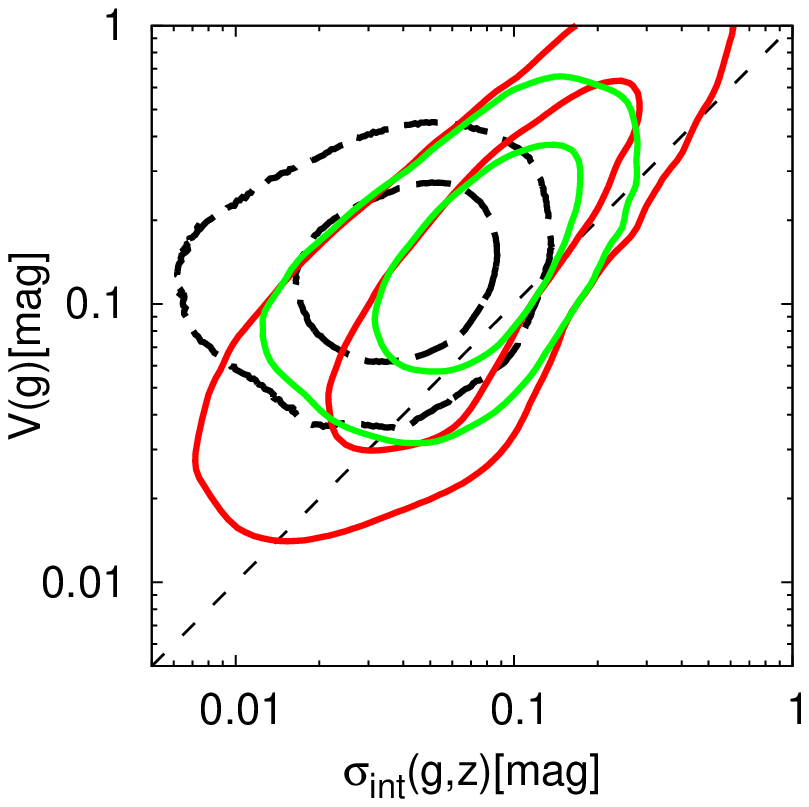}
\hspace{-0.4cm}
\includegraphics[clip, width=2.3in]{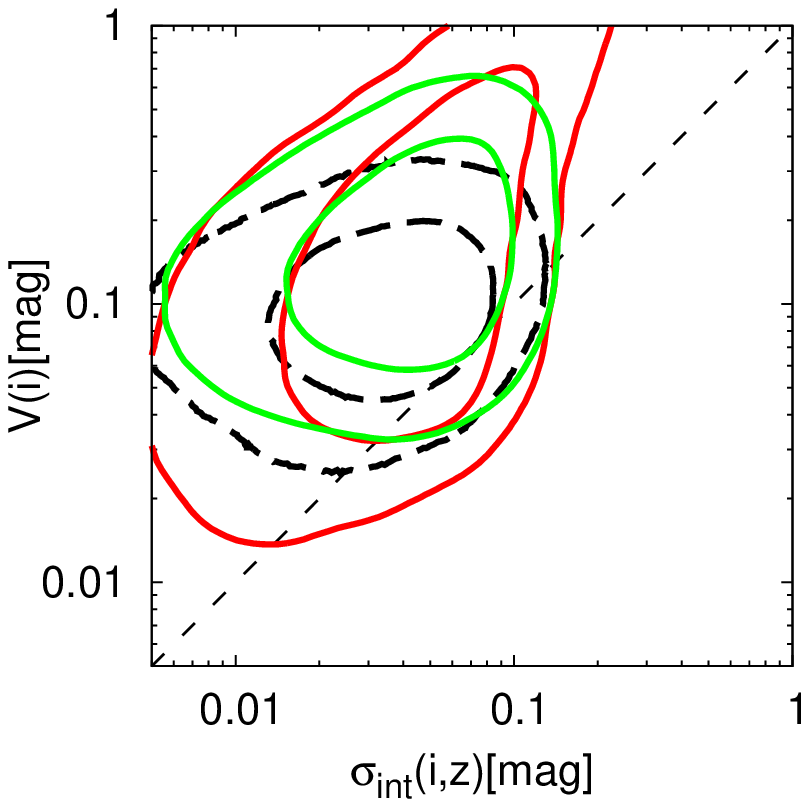}
\hspace{-0.4cm}
}
\leftline{
\includegraphics[clip, width=2.3in]{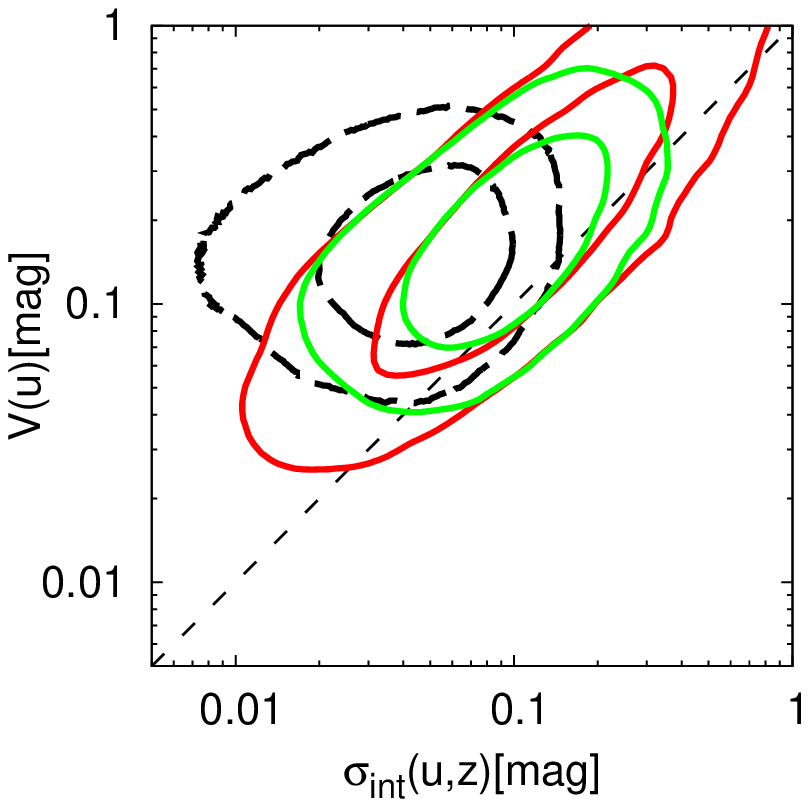}
\hspace{-0.4cm}
}
\caption{
Intrinsic scatter $\sigma_{\text{int}}$ versus variability amplitude $V$ for the Dexter \& Agol inhomogeneous accretion disc model and the SDSS Stripe~82 data for all of the band pairs. 
The solid red lines are the 1$\sigma$ (68.2 per cent) and the 2$\sigma$ (95.4 per cent) contours of the model $\sigma_{\text{int}}$-$V$ relation for the whole of the parameter
space: 0.1 $\leq$ $\sigma_T$ $\leq$ 0.6 (in 0.1 increments), 16 $\leq$ $n$ $\leq$ 4096, $T^*$ = 50 000, 100 000 and 200 000 K, and the specific redshift range for each of the band pairs.
The solid green lines are the same as the red lines but for the restricted parameter space comparable with Dexter \& Agol’s constraints on $\sigma_{T}$ and $n$: 0.35 $\leq$ $\sigma_T$ $\leq$ 0.50 (in 0.05 increments) and 144 $\leq$ $n$ $\leq$ 1024.
The black dashed lines are the same as Fig.~\ref{fig:data_locus}; i.e., the 1$\sigma$ and the 2$\sigma$ contours of the $\sigma_{\text{int}}$-$V$ relation of the SDSS Stripe~82 data. 
A thin dashed straight line in each of the panels indicates $\sigma_{\text{int}} = V$.
}
\label{fig:model_locus}
\end{figure*}

Fig.~\ref{fig:model_locus} shows the intrinsic scatter $\sigma_{\text{int}}$ versus variability amplitude $V$ for the Dexter \& Agol inhomogeneous accretion disc model and the SDSS Stripe~82 data for all of the band pairs, expressed as contours; the solid red lines are the 1$\sigma$ (68.2 per cent, inner dashed line) and the 2$\sigma$ (95.4 per cent, outer dashed line) contours of the model $\sigma_{\text{int}}$-$V$ relation for the whole of the parameter space: 0.1 $\leq$ $\sigma_T$ $\leq$ 0.6 (in 0.1 increments), 16 $\leq$ $n$ $\leq$ 4096, $T^{*}$ $=$ 50 000, 100 000 and 200 000 K, and the specific redshift range for each of the band pairs.
The solid green lines are the same as the red lines but for the restricted parameter space comparable with Dexter \& Agol's constraints on $\sigma_{T}$ and $n$: 0.35 $\leq$ $\sigma_T$ $\leq$ 0.50 (in 0.05 increments) and 144 $\leq$ $n$ $\leq$ 1024.
The black dashed lines are the same as Fig.~\ref{fig:data_locus}; i.e., the 1$\sigma$ and the 2$\sigma$ contours of the $\sigma_{\text{int}}$-$V$ relation of the SDSS Stripe~82 data.

As expected, the inhomogeneous accretion disc model predicts that a tighter magnitude$-$magnitude correlation (i.e., smaller values of $\sigma_{\text{int}}$) can be achieved if we take band pairs with shorter separations of the wavelengths (see $\Delta \lambda_{\text{rest}}$ listed in Table~\ref{tbl:data_properties}), namely $u$-$g$, $g$-$r$, $r$-$i$ and $i$-$z$. 
Actually, if we only focus on the band pairs $u$-$g$, $g$-$r$, $r$-$i$ and $i$-$z$, the model $\sigma_{\text{int}}$-$V$ relation seems to be consistent with the observed $\sigma_{\text{int}}$-$V$ relation. However, we should note the discrepancy in $\sigma_{\text{int}}$ between the model and the data gets clearer for band pairs with longer separations of the effective wavelengths; in particular, for the $u$-$i$, $u$-$z$ and $g$-$z$ band pairs, the model-predicted regions cover only a part of the data regions, and the data generally show smaller values of $\sigma_{\text{int}}$ than those of the model for comparable values of $V$.

To see the discrepancy more clearly, in Fig.~\ref{fig:difference} we show the 1$\sigma$ ranges of $V$ and $\sigma_{\text{int}}$ for the SDSS Stripe~82 data and for the Dexter \& Agol inhomogeneous accretion disc model as a function of the filter combination, obtained by projecting the two-dimensional distributions of $\sigma_{\text{int}}$ and $V$ (shown in Fig.~\ref{fig:model_locus}) on to the $V$- and $\sigma_{\text{int}}$-axes, respectively.
Since in Fig.~\ref{fig:model_locus} and the upper panel of Fig.~\ref{fig:difference} we can see that the variability amplitudes $V$ of the SDSS Stripe~82 quasars are well reproduced by the inhomogeneous accretion disc model with restricted model parameter space comparable with Dexter \& Agol's constraints [this verifies the consistency between model calculations by us and \cite{dex11}], in Fig.~\ref{fig:difference} we show only the results of the model calculations with the restricted model parameters.
As mentioned above, we can clearly see in Fig.~\ref{fig:difference} that the difference for $\sigma_{\text{int}}$ between the data and the model predictions becomes larger when the wavelength separation becomes larger, and the data show smaller values of $\sigma_{\text{int}}$ than those of the model. 
For the band pairs $u$-$i$, $u$-$z$ and $g$-$z$, $\sigma_{\text{int}}$(model) $>$ $\sigma_{\text{int}}$(data) is satisfied for about half of the SDSS Stripe~82 quasars, and this discrepancy cannot be reduced by simply adjusting the model parameters in the Dexter \& Agol inhomogeneous accretion disc model.
A Kolmogorov$-$Smirnov test reveals that the difference between the data and the model distributions of $\sigma_{\text{int}}$ for the $u$-$z$ band pair is statistically significant, and confirms that the null hypothesis (the data are drawn from the model distribution) is rejected even at the level of significance $\alpha = 0.001$.
It should be noted that, as mentioned in Section~\ref{method}, $\sigma_{\text{int}}$(data) evaluated by the broad-band quasar light curves is contaminated by the flux variation of the broad emission lines and the Balmer continuum emission, thus the true values of $\sigma_{\text{int}}$ (i.e., the intrinsic scatter of the magnitude$-$magnitude correlation of the pure accretion disc continuum emission) must be much smaller than those observed.

These results verify the intuition mentioned in Section~\ref{intro};
i.e., Dexter \& Agol's inhomogeneous accretion disc model cannot explain the coherent flux variation within the UV-optical wavelength
range often observed in quasar light curves.

\section{Discussion and Conclusions}
\label{discussion}

\begin{figure}
\center{
\includegraphics[clip, width=3.2in]{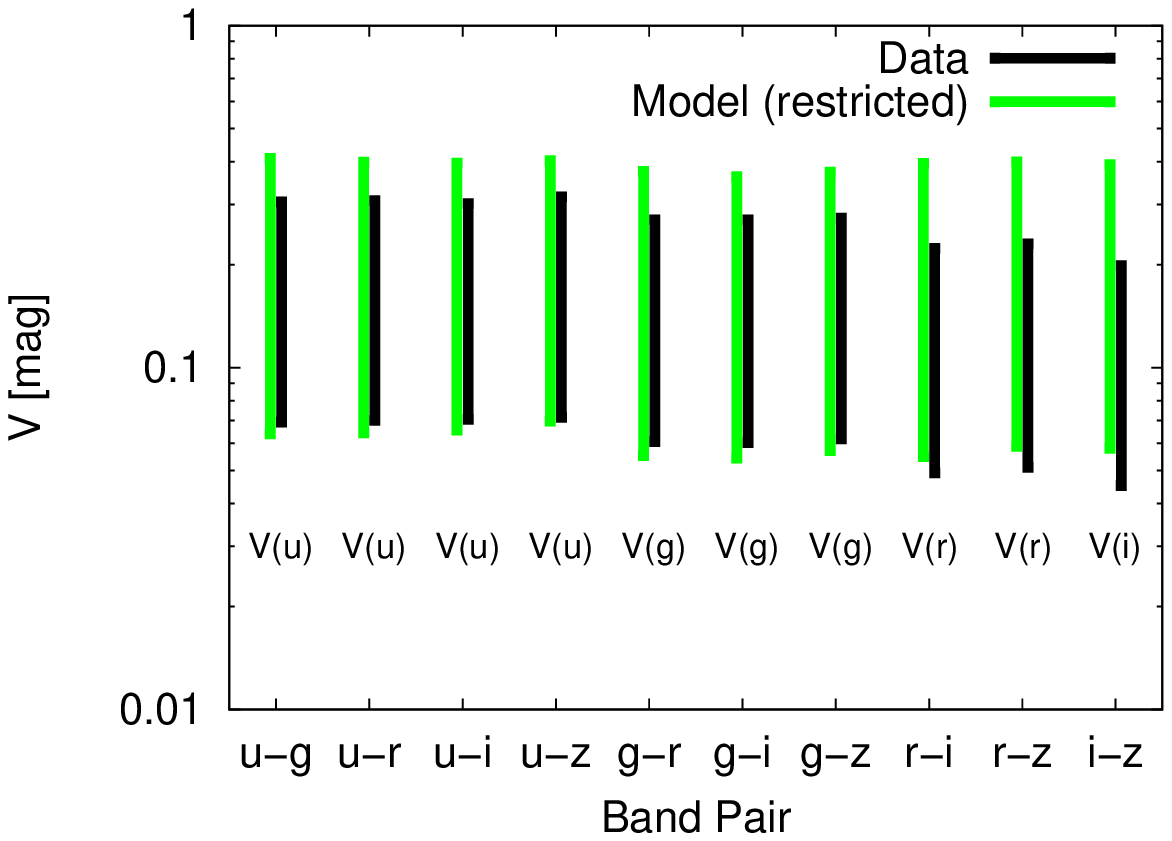}
}
\center{
\includegraphics[clip, width=3.2in]{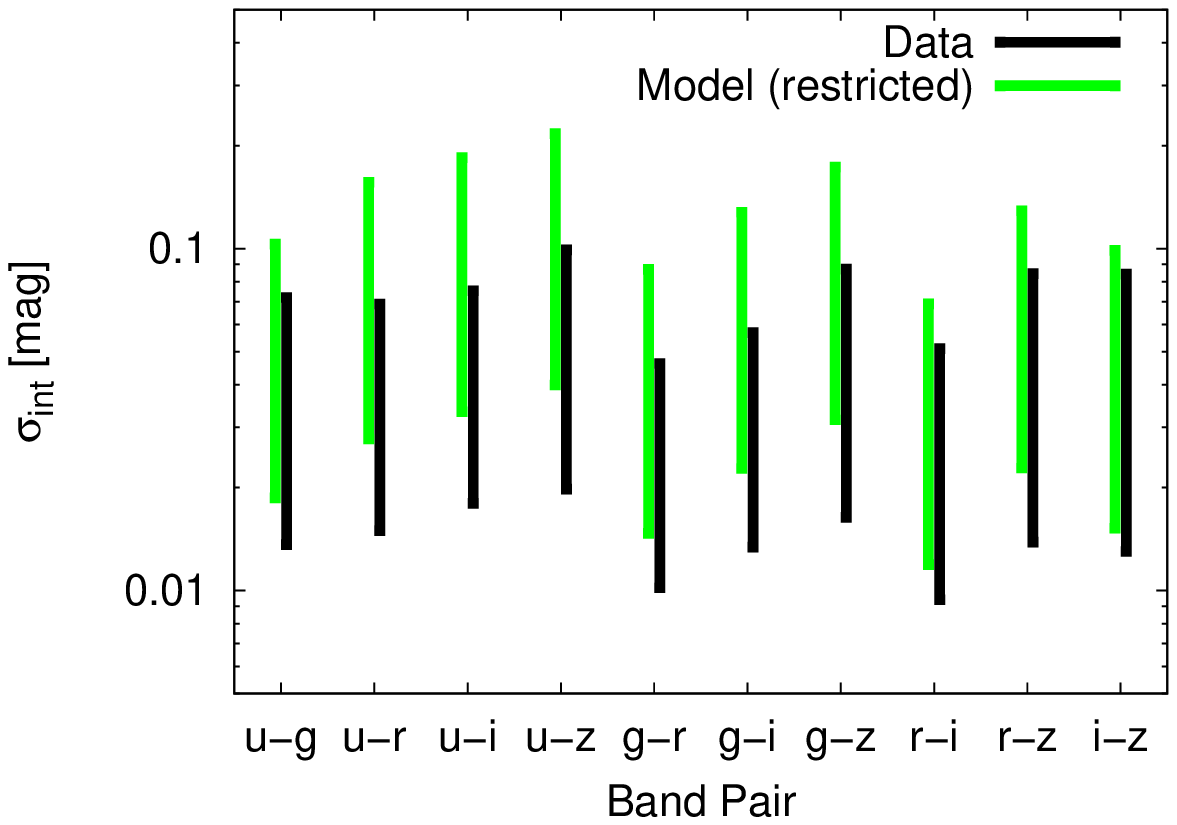}
}
\caption{
$1\sigma$ ranges of $V$ (upper panel) and $\sigma_{\text{int}}$ (lower panel) for the SDSS Stripe~82 data (black) and for the Dexter \& Agol inhomogeneous accretion disc model with the restricted parameter space (green) as a function of the filter combination, obtained by projecting the two-dimensional distributions of $\sigma_{\text{int}}$ and $V$ (shown in Fig.~\ref{fig:model_locus}) on to the $V$- and $\sigma_{\text{int}}$-axes, respectively.
}
\label{fig:difference}
\end{figure}

In this work, we have examined the validity of the inhomogeneous accretion disc model presented by \cite{dex11} in the light of the quasar UV-optical spectral variability. 
As assumed in the Dexter \& Agol inhomogeneous accretion disc model, because the quasar accretion discs have a large physical size, the several years' temperature fluctuations occurring at different radii of the quasar accretion discs must be causally unconnected. 
Observationally, this property of the inhomogeneous accretion disc model ought to result in a weak inter-band magnitude$-$magnitude correlation within the UV-optical wavelength range \citep[see, e.g.,][]{per06,gas11}.
We have used a large sample of long-term (approximately several years), simultaneous five-band light curves of quasars in the SDSS Stripe~82 region to evaluate intrinsic scatter $\sigma_{\text{int}}$(data) in magnitude$-$magnitude space, and compared $\sigma_{\text{int}}$(data) with the Dexter \& Agol inhomogeneous accretion disc model predictions of the intrinsic scatter $\sigma_{\text{int}}$(model). 
We have shown that, especially for the two-band light curves for the $u$-$i$, $u$-$z$ and $g$-$z$ band pairs, $\sigma_{\text{int}}$(model) is generally larger than $\sigma_{\text{int}}$(data) for comparable values of variability amplitude $V$ (Fig.~\ref{fig:model_locus} and Fig.~\ref{fig:difference}), which indicates that the Dexter \& Agol inhomogeneous accretion disc model cannot explain the tight inter-band correlation often observed in the quasar light curves within the UV-optical wavelength range. 
This result indicates that the local temperature fluctuations in the accretion discs are not the main driver of the several years' UV-optical variability of quasars, and the assumption of a strongly inhomogeneous accretion disc is not preferred from the viewpoint of UV-optical spectral variability.

This conclusion seems to be inconsistent with that obtained by \cite{rua14}, in the sense that they concluded that the Dexter \& Agol inhomogeneous accretion disc model accounted well for the spectral variability of SDSS quasars. 
However, as we have already mentioned in Section~\ref{intro}, the difference between our result and that of \cite{rua14} is the expected one: \cite{rua14} compared a geometric mean composite difference spectrum of the SDSS quasars with geometric mean composite difference model spectra, thus the model-predicted incoherent inter-band flux variations were smeared out. 
Because in our analyses the coherence of inter-band flux variations for each individual quasar is compared with each individual time series realization of the model, we can confidently confirm that it is difficult to explain the coherent flux variation within the UV-optical wavelength range in each individual quasar by the Dexter \& Agol inhomogeneous accretion disc model.

Moreover, it should be noted that employing unexpectedly higher-temperature flares than considered here cannot settle the problem. 
If the blackbody flares always have a very high temperature ($\gg$ 100 000 K), the resulting UV-optical spectral variability would show a strong inter-band correlation because the Rayleigh$-$Jeans tail (i.e., a power-law spectrum with $\alpha_{\nu} = 2$) would dominate the whole of the UV-optical wavelength range. 
However, it is known that the observed colour of the quasar variable components is, generally, significantly redder than that of the Rayleigh$-$Jeans spectrum \citep{per06,kok14,rua14}, which indicates that high-temperature blackbody flares also cannot be the main driver of the quasar UV-optical variability.

Our study clarifies the importance of exploring the variability models, which can simultaneously explain the general properties of the several years' AGN variability, namely, the large variability amplitude and the tight inter-band correlation within the UV-optical wavelength range. 
One of the possibilities is the reprocessing model, which assumes the AGN UV-optical variability is caused by reprocessing of X-ray or extreme UV emission \citep[e.g.,][]{kro91}.
Since the fluctuations of this higher-energy photon flux, originating from the innermost region of the accretion disc, propagate from inner to outer radii in the light-crossing time, the reprocessing model can explain the several years' tight inter-band correlation within the UV-optical wavelength range \citep[see][for review]{utt14}.

The reprocessing model predicts that the flux variation occurs at the UV wavelengths prior to the optical wavelengths with a time lag of a few hours (depending on the size of the accretion disc) \citep[e.g.,][]{col99,cac07}.
On the other hand, mass accretion fluctuations propagate from outer to inner radii in the accretion disc, which means that the flux variations caused by the mass accretion fluctuations occur at the optical (longer) wavelengths prior to the UV (shorter) wavelengths \citep[e.g.,][]{are08,utt14}.
Within the UV-optical wavelength range, the inter-band time delay of the AGN continuum for several AGNs are reported to be as expected in the reprocessing model \citep[e.g.,][]{col99,ser05,sug06,cze07,cac07,goi12,che13,loh14,sha14,mch14,ede15}, although results from different observing campaigns are conflicting \citep[see, e.g.,][]{nan98,utt06}.
Moreover, several authors have pointed out that the X-ray luminosity of AGNs is insufficient to power the large UV-optical variability amplitude \citep[e.g.,][]{ant96,gas08,ede14,utt14}.
Obviously, further observational studies are needed to test these models.

Finally, we should note that in this work we only focus on the long-term (i.e., several years) spectral variability of quasars. 
It is possible that the long-term and the short-term (several days) variabilities are driven by totally different physical mechanisms; there remains a possibility that the short-term AGN variability (with small variability amplitude) may be caused by some kind of local activity in AGN accretion discs \citep[e.g.,][and references therein]{col01,cze08,voe11,ai13,cho14,gra14,sun14,mch14,ede14}.
To clarify the true nature of the AGN accretion discs, it is crucial to obtain the multi-epoch dense-time sampling and multi-wavelength light curve data for a large sample of AGN/quasars. 
Ongoing and future wide and deep time-domain surveys, such as the Palomar Transient Factory, the Panoramic Survey Telescope \& Rapid Response System, the Dark Energy Survey, SDSS-IV Time-Domain Spectroscopic Survey and Large Synoptic Survey Telescope \citep{ive08}, will motivate the multi-epoch and multi-wavelength monitoring campaigns for AGN/quasars. 
These observations will clarify the nature of the short- and long-term X-ray-UV-optical-near-infrared correlations, and consequently, of the AGN/quasar accretion disc physics.

\section*{Acknowledgments}

We gratefully acknowledge the referee for his/her constructive advice to improve this paper.
We thank Jason Dexter for help with understanding the inhomogeneous accretion disc model. We also thank Takeo Minezaki and Mamoru Doi for comments and discussions.

The data analysis were in part carried out on common use data analysis computer system (pc06, IDL~8.1) at the Astronomy Data Centre , ADC, of the National Astronomical Observatory of Japan.

Funding for SDSS and SDSS-II has been provided by the Alfred P. Sloan Foundation, the Participating Institutions, the National Science Foundation, the U.S. Department of Energy, the National Aeronautics and Space Administration, the Japanese Monbukagakusho, the Max Planck Society, and the Higher Education Funding Council for England. The SDSS website is {http://www.sdss.org/}.

SDSS is managed by the Astrophysical Research Consortium for the Participating Institutions, which are the American Museum of Natural History, Astrophysical Institute Potsdam, University of Basel, University of Cambridge, Case Western Reserve University, University of Chicago, Drexel University, Fermilab, the Institute for Advanced Study, the Japan Participation Group, Johns Hopkins University, the Joint Institute for Nuclear Astrophysics, the Kavli Institute for Particle Astrophysics and Cosmology, the Korean Scientist Group, the Chinese Academy of Sciences (LAMOST), Los Alamos National Laboratory, the Max-Planck-Institute for Astronomy (MPIA), the Max-Planck-Institute for Astrophysics (MPA), New Mexico State University, Ohio State University, University of Pittsburgh, University of Portsmouth, Princeton University, the United States Naval Observatory, and the University of Washington.

\bibliography{./id}

\begin{thebibliography}{128}
\expandafter\ifx\csname natexlab\endcsname\relax\def\natexlab#1{#1}\fi

\bibitem[{{Abazajian} {et~al}\mbox{.}(2009){Abazajian}, {Adelman-McCarthy},
  {Ag{\"u}eros}, {Allam}, {Allende Prieto}, {An}, {Anderson}, {Anderson},
  {Annis}, {Bahcall}, \& et~al.}]{aba09}
{Abazajian} K.~N. {et~al.}, 2009, \apjs, 182, 543

\bibitem[{{Adelman-McCarthy} {et~al}\mbox{.}(2007){Adelman-McCarthy},
  {Ag{\"u}eros}, {Allam}, {Anderson}, {Anderson}, {Annis}, {Bahcall},
  {Bailer-Jones}, {Baldry}, {Barentine}, {Beers}, {Belokurov}, {Berlind},
  {Bernardi}, {Blanton}, {Bochanski}, {Boroski}, {Bramich}, {Brewington},
  {Brinchmann}, {Brinkmann}, {Brunner}, {Budav{\'a}ri}, {Carey}, {Carliles},
  {Carr}, {Castander}, {Connolly}, {Cool}, {Cunha}, {Csabai}, {Dalcanton},
  {Doi}, {Eisenstein}, {Evans}, {Evans}, {Fan}, {Finkbeiner}, {Friedman},
  {Frieman}, {Fukugita}, {Gillespie}, {Gilmore}, {Glazebrook}, {Gray},
  {Grebel}, {Gunn}, {de Haas}, {Hall}, {Harvanek}, {Hawley}, {Hayes},
  {Heckman}, {Hendry}, {Hennessy}, {Hindsley}, {Hirata}, {Hogan}, {Hogg},
  {Holtzman}, {Ichikawa}, {Ichikawa}, {Ivezi{\'c}}, {Jester}, {Johnston},
  {Jorgensen}, {Juri{\'c}}, {Kauffmann}, {Kent}, {Kleinman}, {Knapp},
  {Kniazev}, {Kron}, {Krzesinski}, {Kuropatkin}, {Lamb}, {Lampeitl}, {Lee},
  {Leger}, {Lima}, {Lin}, {Long}, {Loveday}, {Lupton}, {Mandelbaum}, {Margon},
  {Mart{\'{\i}}nez-Delgado}, {Matsubara}, {McGehee}, {McKay}, {Meiksin},
  {Munn}, {Nakajima}, {Nash}, {Neilsen}, {Newberg}, {Nichol},
  {Nieto-Santisteban}, {Nitta}, {Oyaizu}, {Okamura}, {Ostriker}, {Padmanabhan},
  {Park}, {Peoples}, {Pier}, {Pope}, {Pourbaix}, {Quinn}, {Raddick}, {Re
  Fiorentin}, {Richards}, {Richmond}, {Rix}, {Rockosi}, {Schlegel},
  {Schneider}, {Scranton}, {Seljak}, {Sheldon}, {Shimasaku}, {Silvestri},
  {Smith}, {Smol{\v c}i{\'c}}, {Snedden}, {Stebbins}, {Stoughton}, {Strauss},
  {SubbaRao}, {Suto}, {Szalay}, {Szapudi}, {Szkody}, {Tegmark}, {Thakar},
  {Tremonti}, {Tucker}, {Uomoto}, {Vanden Berk}, {Vandenberg}, {Vidrih},
  {Vogeley}, {Voges}, {Vogt}, {Weinberg}, {West}, {White}, {Wilhite}, {Yanny},
  {Yocum}, {York}, {Zehavi}, {Zibetti}, \& {Zucker}}]{ade07}
{Adelman-McCarthy} J.~K. {et~al.}, 2007, \apjs, 172, 634

\bibitem[{{Ai} {et~al}\mbox{.}(2013){Ai}, {Yuan}, {Zhou}, {Wang}, {Dong},
  {Wang}, \& {Lu}}]{ai13}
{Ai} Y.~L., {Yuan} W., {Zhou} H., {Wang} T.~G., {Dong} X.-B., {Wang} J.~G.,
  {Lu} H.~L., 2013, \aj, 145, 90

\bibitem[{{Ai} {et~al}\mbox{.}(2010){Ai}, {Yuan}, {Zhou}, {Wang}, {Dong},
  {Wang}, \& {Lu}}]{ai10}
{Ai} Y.~L., {Yuan} W., {Zhou} H.~Y., {Wang} T.~G., {Dong} X.-B., {Wang} J.~G.,
  {Lu} H.~L., 2010, \apjl, 716, L31

\bibitem[{{Andrae}, {Kim} \& {Bailer-Jones}(2013){Andrae}, {Kim}, \&
  {Bailer-Jones}}]{and13}
{Andrae} R., {Kim} D.-W., {Bailer-Jones} C.~A.~L., 2013, \aap, 554, A137

\bibitem[{{Antonucci} {et~al}\mbox{.}(1996){Antonucci}, {Geller}, {Goodrich},
  \& {Miller}}]{ant96}
{Antonucci} R., {Geller} R., {Goodrich} R.~W., {Miller} J.~S., 1996, \apj, 472,
  502

\bibitem[{{Aretxaga}, {Cid Fernandes} \& {Terlevich}(1997){Aretxaga}, {Cid
  Fernandes}, \& {Terlevich}}]{are97}
{Aretxaga} I., {Cid Fernandes} R., {Terlevich} R.~J., 1997, \mnras, 286, 271

\bibitem[{{Aretxaga} \& {Terlevich}(1994)}]{are94}
{Aretxaga} I., {Terlevich} R., 1994, \mnras, 269, 462

\bibitem[{{Ar{\'e}valo} {et~al}\mbox{.}(2008){Ar{\'e}valo}, {Uttley}, {Kaspi},
  {Breedt}, {Lira}, \& {McHardy}}]{are08}
{Ar{\'e}valo} P., {Uttley} P., {Kaspi} S., {Breedt} E., {Lira} P., {McHardy}
  I.~M., 2008, \mnras, 389, 1479

\bibitem[{{Baldwin} {et~al}\mbox{.}(1995){Baldwin}, {Ferland}, {Korista}, \&
  {Verner}}]{bal95}
{Baldwin} J., {Ferland} G., {Korista} K., {Verner} D., 1995, \apjl, 455, L119

\bibitem[{{Bauer} {et~al}\mbox{.}(2009){Bauer}, {Baltay}, {Coppi}, {Ellman},
  {Jerke}, {Rabinowitz}, \& {Scalzo}}]{bau09}
{Bauer} A., {Baltay} C., {Coppi} P., {Ellman} N., {Jerke} J., {Rabinowitz} D.,
  {Scalzo} R., 2009, \apj, 696, 1241

\bibitem[{{Blandford} \& {McKee}(1982)}]{bla82}
{Blandford} R.~D., {McKee} C.~F., 1982, \apj, 255, 419

\bibitem[{{Butler} \& {Bloom}(2011)}]{but11}
{Butler} N.~R., {Bloom} J.~S., 2011, \aj, 141, 93

\bibitem[{{Cackett}, {Horne} \& {Winkler}(2007){Cackett}, {Horne}, \&
  {Winkler}}]{cac07}
{Cackett} E.~M., {Horne} K., {Winkler} H., 2007, \mnras, 380, 669

\bibitem[{{Chelouche}(2013)}]{che13}
{Chelouche} D., 2013, \apj, 772, 9

\bibitem[{{Chelouche} {et~al}\mbox{.}(2014){Chelouche}, {Shemmer}, {Cotlier},
  {Barth}, \& {Rafter}}]{che14}
{Chelouche} D., {Shemmer} O., {Cotlier} G.~I., {Barth} A.~J., {Rafter} S.~E.,
  2014, \apj, 785, 140

\bibitem[{{Choi} {et~al}\mbox{.}(2014){Choi}, {Gibson}, {Becker}, {Ivezi{\'c}},
  {Connolly}, {MacLeod}, {Ruan}, \& {Anderson}}]{cho14}
{Choi} Y., {Gibson} R.~R., {Becker} A.~C., {Ivezi{\'c}} {\v Z}., {Connolly}
  A.~J., {MacLeod} C.~L., {Ruan} J.~J., {Anderson} S.~F., 2014, \apj, 782, 37

\bibitem[{{Choloniewski}(1981)}]{cho81}
{Choloniewski} J., 1981, \actaa, 31, 293

\bibitem[{{Cid Fernandes}, {Sodr{\'e}} \& {Vieira da Silva}(2000){Cid
  Fernandes}, {Sodr{\'e}}, \& {Vieira da Silva}}]{cid00}
{Cid Fernandes} R., {Sodr{\'e}}, Jr. L., {Vieira da Silva}, Jr. L., 2000, \apj,
  544, 123

\bibitem[{{Collier} {et~al}\mbox{.}(1999){Collier}, {Horne}, {Wanders}, \&
  {Peterson}}]{col99}
{Collier} S., {Horne} K., {Wanders} I., {Peterson} B.~M., 1999, \mnras, 302,
  L24

\bibitem[{{Collier} \& {Peterson}(2001)}]{col01}
{Collier} S., {Peterson} B.~M., 2001, \apj, 555, 775

\bibitem[{{Courvoisier} \& {Clavel}(1991)}]{cou91}
{Courvoisier} T.~J.-L., {Clavel} J., 1991, \aap, 248, 389

\bibitem[{{Cristiani} {et~al}\mbox{.}(1997){Cristiani}, {Trentini}, {La
  Franca}, \& {Andreani}}]{cri97}
{Cristiani} S., {Trentini} S., {La Franca} F., {Andreani} P., 1997, \aap, 321,
  123

\bibitem[{{Cutri} {et~al}\mbox{.}(1985){Cutri}, {Wisniewski}, {Rieke}, \&
  {Lebofsky}}]{cut85}
{Cutri} R.~M., {Wisniewski} W.~Z., {Rieke} G.~H., {Lebofsky} M.~J., 1985, \apj,
  296, 423

\bibitem[{{Czerny} {et~al}\mbox{.}(2013){Czerny}, {Hryniewicz}, {Maity},
  {Schwarzenberg-Czerny}, {{\.Z}ycki}, \& {Bilicki}}]{cze13}
{Czerny} B., {Hryniewicz} K., {Maity} I., {Schwarzenberg-Czerny} A.,
  {{\.Z}ycki} P.~T., {Bilicki} M., 2013, \aap, 556, A97

\bibitem[{{Czerny} \& {Janiuk}(2007)}]{cze07}
{Czerny} B., {Janiuk} A., 2007, \aap, 464, 167

\bibitem[{{Czerny} {et~al}\mbox{.}(2008){Czerny}, {Siemiginowska}, {Janiuk}, \&
  {Gupta}}]{cze08}
{Czerny} B., {Siemiginowska} A., {Janiuk} A., {Gupta} A.~C., 2008, \mnras, 386,
  1557

\bibitem[{{D'Agostini}(2005)}]{dag05}
{D'Agostini} G., 2005, ArXiv Physics e-prints: physics/0511182

\bibitem[{{Dai} {et~al}\mbox{.}(2010){Dai}, {Kochanek}, {Chartas},
  {Koz{\l}owski}, {Morgan}, {Garmire}, \& {Agol}}]{dai10}
{Dai} X., {Kochanek} C.~S., {Chartas} G., {Koz{\l}owski} S., {Morgan} C.~W.,
  {Garmire} G., {Agol} E., 2010, \apj, 709, 278

\bibitem[{{de Vries} {et~al}\mbox{.}(2005){de Vries}, {Becker}, {White}, \&
  {Loomis}}]{dev05}
{de Vries} W.~H., {Becker} R.~H., {White} R.~L., {Loomis} C., 2005, \aj, 129,
  615

\bibitem[{{Dexter} \& {Agol}(2011)}]{dex11}
{Dexter} J., {Agol} E., 2011, \apjl, 727, L24

\bibitem[{{di Clemente} {et~al}\mbox{.}(1996){di Clemente}, {Giallongo},
  {Natali}, {Trevese}, \& {Vagnetti}}]{dic96}
{di Clemente} A., {Giallongo} E., {Natali} G., {Trevese} D., {Vagnetti} F.,
  1996, \apj, 463, 466

\bibitem[{{Edelson} {et~al}\mbox{.}(2015){Edelson}, {Gelbord}, {Horne},
  {McHardy}, {Peterson}, {Arevalo}, {Breeveld}, {DeRosa}, {Evans}, {Goad},
  {Kriss}, {Brandt}, {Gehrels}, {Grupe}, {Kennea}, {Kochanek}, {Nousek},
  {Papadakis}, {Siegel}, {Starkey}, {Uttley}, {Vaughan}, {Young}, {Barth},
  {Bentz}, {Brewer}, {Crenshaw}, {Dalla Bonta}, {De Lorenzo-Caceres}, {Denney},
  {Dietrich}, {Ely}, {Fausnaugh}, {Grier}, {Hall}, {Kaastra}, {Kelly},
  {Korista}, {Lira}, {Mathur}, {Netzer}, {Pancoast}, {Pei}, {Pogge},
  {Schimoia}, {Treu}, {Vestergaard}, {Villforth}, {Yan}, \& {Zu}}]{ede15}
{Edelson} R. {et~al.}, 2015, ArXiv e-prints: 1501.05951

\bibitem[{{Edelson} {et~al}\mbox{.}(2014){Edelson}, {Vaughan}, {Malkan},
  {Kelly}, {Smith}, {Boyd}, \& {Mushotzky}}]{ede14}
{Edelson} R., {Vaughan} S., {Malkan} M., {Kelly} B.~C., {Smith} K.~L., {Boyd}
  P.~T., {Mushotzky} R., 2014, \apj, 795, 2

\bibitem[{{Feigelson} \& {Jogesh Babu}(2012)}]{fei12}
{Feigelson} E.~D., {Jogesh Babu} G., 2012, {Modern Statistical Methods for
  Astronomy}

\bibitem[{{Frank}, {King} \& {Raine}(1992){Frank}, {King}, \& {Raine}}]{fra92}
{Frank} J., {King} A., {Raine} D., 1992, {Accretion power in astrophysics.}

\bibitem[{{Frieman} {et~al}\mbox{.}(2008){Frieman}, {Bassett}, {Becker},
  {Choi}, {Cinabro}, {DeJongh}, {Depoy}, {Dilday}, {Doi}, {Garnavich}, {Hogan},
  {Holtzman}, {Im}, {Jha}, {Kessler}, {Konishi}, {Lampeitl}, {Marriner},
  {Marshall}, {McGinnis}, {Miknaitis}, {Nichol}, {Prieto}, {Riess}, {Richmond},
  {Romani}, {Sako}, {Schneider}, {Smith}, {Takanashi}, {Tokita}, {van der
  Heyden}, {Yasuda}, {Zheng}, {Adelman-McCarthy}, {Annis}, {Assef},
  {Barentine}, {Bender}, {Blandford}, {Boroski}, {Bremer}, {Brewington},
  {Collins}, {Crotts}, {Dembicky}, {Eastman}, {Edge}, {Edmondson}, {Elson},
  {Eyler}, {Filippenko}, {Foley}, {Frank}, {Goobar}, {Gueth}, {Gunn},
  {Harvanek}, {Hopp}, {Ihara}, {Ivezi{\'c}}, {Kahn}, {Kaplan}, {Kent},
  {Ketzeback}, {Kleinman}, {Kollatschny}, {Kron}, {Krzesi{\'n}ski}, {Lamenti},
  {Leloudas}, {Lin}, {Long}, {Lucey}, {Lupton}, {Malanushenko}, {Malanushenko},
  {McMillan}, {Mendez}, {Morgan}, {Morokuma}, {Nitta}, {Ostman}, {Pan},
  {Rockosi}, {Romer}, {Ruiz-Lapuente}, {Saurage}, {Schlesinger}, {Snedden},
  {Sollerman}, {Stoughton}, {Stritzinger}, {Subba Rao}, {Tucker}, {Vaisanen},
  {Watson}, {Watters}, {Wheeler}, {Yanny}, \& {York}}]{fri08}
{Frieman} J.~A. {et~al.}, 2008, \aj, 135, 338

\bibitem[{{Gallastegui-Aizpun} \& {Sarajedini}(2014)}]{gal14}
{Gallastegui-Aizpun} U., {Sarajedini} V.~L., 2014, \mnras, 444, 3078

\bibitem[{{Gaskell}(2008)}]{gas08}
{Gaskell} C.~M., 2008, in Revista Mexicana de Astronomia y Astrofisica
  Conference Series, Vol.~32, Revista Mexicana de Astronomia y Astrofisica
  Conference Series, pp. 1--11

\bibitem[{{Gaskell}(2011)}]{gas11}
{Gaskell} C.~M., 2011, Baltic Astronomy, 20, 392

\bibitem[{{Gelman} \& {Rubin}(1992)}]{gel92}
{Gelman} A., {Rubin} D.~B., 1992, Statist. Sci., 7, 457

\bibitem[{{Gezari} {et~al}\mbox{.}(2013){Gezari}, {Martin}, {Forster}, {Neill},
  {Huber}, {Heckman}, {Bianchi}, {Morrissey}, {Neff}, {Seibert},
  {Schiminovich}, {Wyder}, {Burgett}, {Chambers}, {Kaiser}, {Magnier}, {Price},
  \& {Tonry}}]{gez13}
{Gezari} S. {et~al.}, 2013, \apj, 766, 60

\bibitem[{{Giveon} {et~al}\mbox{.}(1999){Giveon}, {Maoz}, {Kaspi}, {Netzer}, \&
  {Smith}}]{giv99}
{Giveon} U., {Maoz} D., {Kaspi} S., {Netzer} H., {Smith} P.~S., 1999, \mnras,
  306, 637

\bibitem[{{Goicoechea} {et~al}\mbox{.}(2012){Goicoechea}, {Shalyapin},
  {Gil-Merino}, \& {Braga}}]{goi12}
{Goicoechea} L.~J., {Shalyapin} V.~N., {Gil-Merino} R., {Braga} V.~F., 2012,
  Journal of Physics Conference Series, 372, 012058

\bibitem[{{Graham} {et~al}\mbox{.}(2014){Graham}, {Djorgovski}, {Drake},
  {Mahabal}, {Chang}, {Stern}, {Donalek}, \& {Glikman}}]{gra14}
{Graham} M.~J., {Djorgovski} S.~G., {Drake} A.~J., {Mahabal} A.~A., {Chang} M.,
  {Stern} D., {Donalek} C., {Glikman} E., 2014, \mnras, 439, 703

\bibitem[{{Gu} \& {Li}(2013)}]{gu13a}
{Gu} M.~F., {Li} S.-L., 2013, \aap, 554, A51

\bibitem[{{Gull}(1989)}]{gul89}
{Gull} S.~F., 1989, in Maximum Entropy and Bayesian Methods, ed. J. Skilling
  (Dordrecht: Kluwer), 511

\bibitem[{{Hagen-Thorn}(1997)}]{hag97}
{Hagen-Thorn} V.~A., 1997, Astronomy Letters, 23, 19

\bibitem[{{Hao} {et~al}\mbox{.}(2010){Hao}, {Elvis}, {Civano}, {Lanzuisi},
  {Brusa}, {Lusso}, {Zamorani}, {Comastri}, {Bongiorno}, {Impey}, {Koekemoer},
  {Le Floc'h}, {Salvato}, {Sanders}, {Trump}, \& {Vignali}}]{hao10}
{Hao} H. {et~al.}, 2010, \apjl, 724, L59

\bibitem[{{Hawkins}(1993)}]{haw93}
{Hawkins} M.~R.~S., 1993, \nat, 366, 242

\bibitem[{{Hawkins}(2002)}]{haw02}
{Hawkins} M.~R.~S., 2002, \mnras, 329, 76

\bibitem[{{Hawkins}(2003)}]{haw03}
{Hawkins} M.~R.~S., 2003, \mnras, 344, 492

\bibitem[{{Hernitschek} {et~al}\mbox{.}(2014){Hernitschek}, {Rix}, {Bovy}, \&
  {Morganson}}]{her14}
{Hernitschek} N., {Rix} H.-W., {Bovy} J., {Morganson} E., 2014, ArXiv e-prints:
  1412.6531

\bibitem[{{Hewett} \& {Wild}(2010)}]{hew10}
{Hewett} P.~C., {Wild} V., 2010, \mnras, 405, 2302

\bibitem[{{Hirose}, {Krolik} \& {Blaes}(2009){Hirose}, {Krolik}, \&
  {Blaes}}]{hir09a}
{Hirose} S., {Krolik} J.~H., {Blaes} O., 2009, \apj, 691, 16

\bibitem[{{Hogg}, {Bovy} \& {Lang}(2010){Hogg}, {Bovy}, \& {Lang}}]{hog10}
{Hogg} D.~W., {Bovy} J., {Lang} D., 2010, ArXiv e-prints: 1008.4686

\bibitem[{{H{\"o}nig} \& {Kishimoto}(2010)}]{hon10}
{H{\"o}nig} S.~F., {Kishimoto} M., 2010, \aap, 523, A27

\bibitem[{{Ivezic} {et~al}\mbox{.}(2008){Ivezic}, {Tyson}, {Abel}, {Acosta},
  {Allsman}, {AlSayyad}, {Anderson}, {Andrew}, {Angel}, {Angeli}, {Ansari},
  {Antilogus}, {Arndt}, {Astier}, {Aubourg}, {Axelrod}, {Bard}, {Barr},
  {Barrau}, {Bartlett}, {Bauman}, {Beaumont}, {Becker}, {Becla}, {Beldica},
  {Bellavia}, {Blanc}, {Blandford}, {Bloom}, {Bogart}, {Borne}, {Bosch},
  {Boutigny}, {Brandt}, {Brown}, {Bullock}, {Burchat}, {Burke}, {Cagnoli},
  {Calabrese}, {Chandrasekharan}, {Chesley}, {Cheu}, {Chiang}, {Claver},
  {Connolly}, {Cook}, {Cooray}, {Covey}, {Cribbs}, {Cui}, {Cutri}, {Daubard},
  {Daues}, {Delgado}, {Digel}, {Doherty}, {Dubois}, {Dubois-Felsmann},
  {Durech}, {Eracleous}, {Ferguson}, {Frank}, {Freemon}, {Gangler}, {Gawiser},
  {Geary}, {Gee}, {Geha}, {Gibson}, {Gilmore}, {Glanzman}, {Goodenow},
  {Gressler}, {Gris}, {Guyonnet}, {Hascall}, {Haupt}, {Hernandez}, {Hogan},
  {Huang}, {Huffer}, {Innes}, {Jacoby}, {Jain}, {Jee}, {Jernigan},
  {Jevremovic}, {Johns}, {Jones}, {Juramy-Gilles}, {Juric}, {Kahn}, {Kalirai},
  {Kallivayalil}, {Kalmbach}, {Kantor}, {Kasliwal}, {Kessler}, {Kirkby},
  {Knox}, {Kotov}, {Krabbendam}, {Krughoff}, {Kubanek}, {Kuczewski},
  {Kulkarni}, {Lambert}, {Le Guillou}, {Levine}, {Liang}, {Lim}, {Lintott},
  {Lupton}, {Mahabal}, {Marshall}, {Marshall}, {May}, {McKercher}, {Migliore},
  {Miller}, {Mills}, {Monet}, {Moniez}, {Neill}, {Nief}, {Nomerotski},
  {Nordby}, {O'Connor}, {Oliver}, {Olivier}, {Olsen}, {Ortiz}, {Owen}, {Pain},
  {Peterson}, {Petry}, {Pierfederici}, {Pietrowicz}, {Pike}, {Pinto}, {Plante},
  {Plate}, {Price}, {Prouza}, {Radeka}, {Rajagopal}, {Rasmussen}, {Regnault},
  {Ridgway}, {Ritz}, {Rosing}, {Roucelle}, {Rumore}, {Russo}, {Saha},
  {Sassolas}, {Schalk}, {Schindler}, {Schneider}, {Schumacher}, {Sebag},
  {Sembroski}, {Seppala}, {Shipsey}, {Silvestri}, {Smith}, {Smith}, {Strauss},
  {Stubbs}, {Sweeney}, {Szalay}, {Takacs}, {Thaler}, {Van Berg}, {Vanden Berk},
  {Vetter}, {Virieux}, {Xin}, {Walkowicz}, {Walter}, {Wang}, {Warner},
  {Willman}, {Wittman}, {Wolff}, {Wood-Vasey}, {Yoachim}, {Zhan}, \& {for the
  LSST Collaboration}}]{ive08}
{Ivezic} Z. {et~al.}, 2008, ArXiv e-prints: 0805.2366

\bibitem[{{Jiang}, {Stone} \& {Davis}(2013){Jiang}, {Stone}, \&
  {Davis}}]{jiang13}
{Jiang} Y.-F., {Stone} J.~M., {Davis} S.~W., 2013, \apj, 767, 148

\bibitem[{{Jim{\'e}nez-Vicente} {et~al}\mbox{.}(2014){Jim{\'e}nez-Vicente},
  {Mediavilla}, {Kochanek}, {Mu{\~n}oz}, {Motta}, {Falco}, \&
  {Mosquera}}]{jim14}
{Jim{\'e}nez-Vicente} J., {Mediavilla} E., {Kochanek} C.~S., {Mu{\~n}oz} J.~A.,
  {Motta} V., {Falco} E., {Mosquera} A.~M., 2014, \apj, 783, 47

\bibitem[{{Kato}, {Fukue} \& {Mineshige}(2008){Kato}, {Fukue}, \&
  {Mineshige}}]{kat08}
{Kato} S., {Fukue} J., {Mineshige} S., 2008, {Black-Hole Accretion Disks ---
  Towards a New Paradigm ---}

\bibitem[{{Kawaguchi} {et~al}\mbox{.}(1998){Kawaguchi}, {Mineshige}, {Umemura},
  \& {Turner}}]{kaw98}
{Kawaguchi} T., {Mineshige} S., {Umemura} M., {Turner} E.~L., 1998, \apj, 504,
  671

\bibitem[{{Kawaguchi}, {Shimura} \& {Mineshige}(2001){Kawaguchi}, {Shimura}, \&
  {Mineshige}}]{kaw01}
{Kawaguchi} T., {Shimura} T., {Mineshige} S., 2001, \apj, 546, 966

\bibitem[{{Kelly}(2007)}]{kel07}
{Kelly} B.~C., 2007, \apj, 665, 1489

\bibitem[{{Kelly}, {Bechtold} \& {Siemiginowska}(2009){Kelly}, {Bechtold}, \&
  {Siemiginowska}}]{kel09}
{Kelly} B.~C., {Bechtold} J., {Siemiginowska} A., 2009, \apj, 698, 895

\bibitem[{{Kishimoto}, {Antonucci} \& {Blaes}(2005){Kishimoto}, {Antonucci}, \&
  {Blaes}}]{kis05}
{Kishimoto} M., {Antonucci} R., {Blaes} O., 2005, \mnras, 364, 640

\bibitem[{{Kishimoto} {et~al}\mbox{.}(2008){Kishimoto}, {Antonucci}, {Blaes},
  {Lawrence}, {Boisson}, {Albrecht}, \& {Leipski}}]{kis08}
{Kishimoto} M., {Antonucci} R., {Blaes} O., {Lawrence} A., {Boisson} C.,
  {Albrecht} M., {Leipski} C., 2008, \nat, 454, 492

\bibitem[{{Kishimoto} {et~al}\mbox{.}(2007){Kishimoto}, {H{\"o}nig}, {Beckert},
  \& {Weigelt}}]{kis07}
{Kishimoto} M., {H{\"o}nig} S.~F., {Beckert} T., {Weigelt} G., 2007, \aap, 476,
  713

\bibitem[{{Kokubo} {et~al}\mbox{.}(2014){Kokubo}, {Morokuma}, {Minezaki},
  {Doi}, {Kawaguchi}, {Sameshima}, \& {Koshida}}]{kok14}
{Kokubo} M., {Morokuma} T., {Minezaki} T., {Doi} M., {Kawaguchi} T.,
  {Sameshima} H., {Koshida} S., 2014, \apj, 783, 46

\bibitem[{{Korista} {et~al}\mbox{.}(1995){Korista}, {Alloin}, {Barr}, {Clavel},
  {Cohen}, {Crenshaw}, {Evans}, {Horne}, {Koratkar}, {Kriss}, {Krolik},
  {Malkan}, {Morris}, {Netzer}, {O'Brien}, {Peterson}, {Reichert},
  {Rodriguez-Pascual}, {Wamsteker}, {Anderson}, {Axon}, {Benitez}, {Berlind},
  {Bertram}, {Blackwell}, {Bochkarev}, {Boisson}, {Carini}, {Carrillo},
  {Carone}, {Cheng}, {Christensen}, {Chuvaev}, {Dietrich}, {Dokter},
  {Doroshenko}, {Dultzin-Hacyan}, {England}, {Espey}, {Filippenko}, {Gaskell},
  {Goad}, {Ho}, {Huchra}, {Jiang}, {Kaspi}, {Kollatschny}, {Laor}, {Luminet},
  {MacAlpine}, {MacKenty}, {Malkov}, {Maoz}, {Martin}, {Matheson}, {McCollum},
  {Merkulova}, {Metik}, {Mignoli}, {Miller}, {Pastoriza}, {Pelat}, {Penfold},
  {Perez}, {Perola}, {Persaud}, {Peters}, {Pitts}, {Pogge}, {Pronik}, {Pronik},
  {Ptak}, {Rawley}, {Recondo-Gonzalez}, {Rodriguez-Espinosa}, {Romanishin},
  {Sadun}, {Salamanca}, {Santos-Lleo}, {Sekiguchi}, {Sergeev}, {Shapovalova},
  {Shields}, {Shrader}, {Shull}, {Silbermann}, {Sitko}, {Skillman}, {Smith},
  {Smith}, {Snijders}, {Sparke}, {Stirpe}, {Stoner}, {Sun}, {Thiele}, {Tokarz},
  {Tsvetanov}, {Turnshek}, {Veilleux}, {Wagner}, {Wagner}, {Wanders}, {Wang},
  {Welsh}, {Weymann}, {White}, {Wilkes}, {Wills}, {Winge}, {Wu}, \&
  {Zou}}]{kor95}
{Korista} K.~T. {et~al.}, 1995, \apjs, 97, 285

\bibitem[{{Korista} \& {Goad}(2001)}]{kor01}
{Korista} K.~T., {Goad} M.~R., 2001, \apj, 553, 695

\bibitem[{{Korista} \& {Goad}(2004)}]{kor04}
{Korista} K.~T., {Goad} M.~R., 2004, \apj, 606, 749

\bibitem[{{Koshida} {et~al}\mbox{.}(2014){Koshida}, {Minezaki}, {Yoshii},
  {Kobayashi}, {Sakata}, {Sugawara}, {Enya}, {Suganuma}, {Tomita}, {Aoki}, \&
  {Peterson}}]{kos14}
{Koshida} S. {et~al.}, 2014, \apj, 788, 159

\bibitem[{{Koz{\l}owski} {et~al}\mbox{.}(2010){Koz{\l}owski}, {Kochanek},
  {Udalski}, {Wyrzykowski}, {Soszy{\'n}ski}, {Szyma{\'n}ski}, {Kubiak},
  {Pietrzy{\'n}ski}, {Szewczyk}, {Ulaczyk}, {Poleski}, \& {OGLE
  Collaboration}}]{koz10}
{Koz{\l}owski} S. {et~al.}, 2010, \apj, 708, 927

\bibitem[{{Krolik} {et~al}\mbox{.}(1991){Krolik}, {Horne}, {Kallman}, {Malkan},
  {Edelson}, \& {Kriss}}]{kro91}
{Krolik} J.~H., {Horne} K., {Kallman} T.~R., {Malkan} M.~A., {Edelson} R.~A.,
  {Kriss} G.~A., 1991, \apj, 371, 541

\bibitem[{{LaMassa} {et~al}\mbox{.}(2014){LaMassa}, {Cales}, {Moran}, {Myers},
  {Richards}, {Eracleous}, {Heckman}, {Gallo}, \& {Urry}}]{lam14}
{LaMassa} S.~M. {et~al.}, 2014, ArXiv e-prints: 1412.2136

\bibitem[{{Li} \& {Cao}(2008)}]{li08}
{Li} S.-L., {Cao} X., 2008, \mnras, 387, L41

\bibitem[{{Lira} {et~al}\mbox{.}(2011){Lira}, {Ar{\'e}valo}, {Uttley},
  {McHardy}, \& {Breedt}}]{lir11}
{Lira} P., {Ar{\'e}valo} P., {Uttley} P., {McHardy} I., {Breedt} E., 2011,
  \mnras, 415, 1290

\bibitem[{{Liu} {et~al}\mbox{.}(2008){Liu}, {Bai}, {Zhao}, \& {Ma}}]{liu08}
{Liu} H.~T., {Bai} J.~M., {Zhao} X.~H., {Ma} L., 2008, \apj, 677, 884

\bibitem[{{Lohfink} {et~al}\mbox{.}(2014){Lohfink}, {Reynolds}, {Vasudevan},
  {Mushotzky}, \& {Miller}}]{loh14}
{Lohfink} A.~M., {Reynolds} C.~S., {Vasudevan} R., {Mushotzky} R.~F., {Miller}
  N.~A., 2014, \apj, 788, 10

\bibitem[{{MacLeod} {et~al}\mbox{.}(2010){MacLeod}, {Ivezi{\'c}}, {Kochanek},
  {Koz{\l}owski}, {Kelly}, {Bullock}, {Kimball}, {Sesar}, {Westman}, {Brooks},
  {Gibson}, {Becker}, \& {de Vries}}]{mac10}
{MacLeod} C.~L. {et~al.}, 2010, \apj, 721, 1014

\bibitem[{{MacLeod} {et~al}\mbox{.}(2012){MacLeod}, {Ivezi{\'c}}, {Sesar}, {de
  Vries}, {Kochanek}, {Kelly}, {Becker}, {Lupton}, {Hall}, {Richards},
  {Anderson}, \& {Schneider}}]{mac12}
{MacLeod} C.~L. {et~al.}, 2012, \apj, 753, 106

\bibitem[{{McHardy} {et~al}\mbox{.}(2014){McHardy}, {Cameron}, {Dwelly},
  {Connolly}, {Lira}, {Emmanoulopoulos}, {Gelbord}, {Breedt}, {Arevalo}, \&
  {Uttley}}]{mch14}
{McHardy} I.~M. {et~al.}, 2014, \mnras, 444, 1469

\bibitem[{{Meusinger}, {Hinze} \& {de Hoon}(2011){Meusinger}, {Hinze}, \& {de
  Hoon}}]{meu11}
{Meusinger} H., {Hinze} A., {de Hoon} A., 2011, \aap, 525, A37

\bibitem[{{Meusinger} \& {Weiss}(2013)}]{meu13}
{Meusinger} H., {Weiss} V., 2013, \aap, 560, A104

\bibitem[{{Morganson} {et~al}\mbox{.}(2014){Morganson}, {Burgett}, {Chambers},
  {Green}, {Kaiser}, {Magnier}, {Marshall}, {Morgan}, {Price}, {Rix},
  {Schlafly}, {Tonry}, \& {Walter}}]{mor14}
{Morganson} E. {et~al.}, 2014, \apj, 784, 92

\bibitem[{{Nandra} {et~al}\mbox{.}(1998){Nandra}, {Clavel}, {Edelson},
  {George}, {Malkan}, {Mushotzky}, {Peterson}, \& {Turner}}]{nan98}
{Nandra} K., {Clavel} J., {Edelson} R.~A., {George} I.~M., {Malkan} M.~A.,
  {Mushotzky} R.~F., {Peterson} B.~M., {Turner} T.~J., 1998, \apj, 505, 594

\bibitem[{{Novikov} \& {Thorne}(1973)}]{nov73}
{Novikov} I.~D., {Thorne} K.~S., 1973, in Black Holes (Les Astres Occlus),
  {Dewitt} C., {Dewitt} B.~S., eds., pp. 343--450

\bibitem[{{O'Brien}, {Goad} \& {Gondhalekar}(1995){O'Brien}, {Goad}, \&
  {Gondhalekar}}]{obr95}
{O'Brien} P.~T., {Goad} M.~R., {Gondhalekar} P.~M., 1995, \mnras, 275, 1125

\bibitem[{{Oknyansky} {et~al}\mbox{.}(2014){Oknyansky}, {Metlova}, {Taranova},
  {Shenavrin}, {Artamonov}, \& {Gaskell}}]{okn14}
{Oknyansky} V.~L., {Metlova} N.~V., {Taranova} O.~G., {Shenavrin} V.~I.,
  {Artamonov} B.~P., {Gaskell} C.~M., 2014, Astronomy Letters, 40, 527

\bibitem[{{Palanque-Delabrouille} {et~al}\mbox{.}(2011){Palanque-Delabrouille},
  {Yeche}, {Myers}, {Petitjean}, {Ross}, {Sheldon}, {Aubourg}, {Delubac}, {Le
  Goff}, {P{\^a}ris}, {Rich}, {Dawson}, {Schneider}, \& {Weaver}}]{pal11}
{Palanque-Delabrouille} N. {et~al.}, 2011, \aap, 530, A122

\bibitem[{{Park} {et~al}\mbox{.}(2012){Park}, {Kelly}, {Woo}, \&
  {Treu}}]{par12}
{Park} D., {Kelly} B.~C., {Woo} J.-H., {Treu} T., 2012, \apjs, 203, 6

\bibitem[{{Pereyra} {et~al}\mbox{.}(2006){Pereyra}, {Vanden Berk}, {Turnshek},
  {Hillier}, {Wilhite}, {Kron}, {Schneider}, \& {Brinkmann}}]{per06}
{Pereyra} N.~A., {Vanden Berk} D.~E., {Turnshek} D.~A., {Hillier} D.~J.,
  {Wilhite} B.~C., {Kron} R.~G., {Schneider} D.~P., {Brinkmann} J., 2006, \apj,
  642, 87

\bibitem[{{Peterson} \& {Horne}(2004)}]{pet04}
{Peterson} B.~M., {Horne} K., 2004, Astronomische Nachrichten, 325, 248

\bibitem[{{Pooley} {et~al}\mbox{.}(2007){Pooley}, {Blackburne}, {Rappaport}, \&
  {Schechter}}]{poo07}
{Pooley} D., {Blackburne} J.~A., {Rappaport} S., {Schechter} P.~L., 2007, \apj,
  661, 19

\bibitem[{{Ruan} {et~al}\mbox{.}(2014){Ruan}, {Anderson}, {Dexter}, \&
  {Agol}}]{rua14}
{Ruan} J.~J., {Anderson} S.~F., {Dexter} J., {Agol} E., 2014, \apj, 783, 105

\bibitem[{{Sakata} {et~al}\mbox{.}(2010){Sakata}, {Minezaki}, {Yoshii},
  {Kobayashi}, {Koshida}, {Aoki}, {Enya}, {Tomita}, {Suganuma}, {Katsuno
  Uchimoto}, \& {Sugawara}}]{sak10}
{Sakata} Y. {et~al.}, 2010, \apj, 711, 461

\bibitem[{{Sakata} {et~al}\mbox{.}(2011){Sakata}, {Morokuma}, {Minezaki},
  {Yoshii}, {Kobayashi}, {Koshida}, \& {Sameshima}}]{sak11}
{Sakata} Y., {Morokuma} T., {Minezaki} T., {Yoshii} Y., {Kobayashi} Y.,
  {Koshida} S., {Sameshima} H., 2011, \apj, 731, 50

\bibitem[{{Sako} {et~al}\mbox{.}(2014){Sako}, {Bassett}, {Becker}, {Brown},
  {Campbell}, {Cane}, {Cinabro}, {D'Andrea}, {Dawson}, {DeJongh}, {Depoy},
  {Dilday}, {Doi}, {Filippenko}, {Fischer}, {Foley}, {Frieman}, {Galbany},
  {Garnavich}, {Goobar}, {Gupta}, {Hill}, {Hayden}, {Hlozek}, {Holtzman},
  {Hopp}, {Jha}, {Kessler}, {Kollatschny}, {Leloudas}, {Marriner}, {Marshall},
  {Miquel}, {Morokuma}, {Mosher}, {Nichol}, {Nordin}, {Olmstead}, {Ostman},
  {Prieto}, {Richmond}, {Romani}, {Sollerman}, {Stritzinger}, {Schneider},
  {Smith}, {Wheeler}, {Yasuda}, \& {Zheng}}]{sak14}
{Sako} M. {et~al.}, 2014, ArXiv e-prints: 1401.3317

\bibitem[{{Schmidt} {et~al}\mbox{.}(2010){Schmidt}, {Marshall}, {Rix},
  {Jester}, {Hennawi}, \& {Dobler}}]{schmidt10}
{Schmidt} K.~B., {Marshall} P.~J., {Rix} H.-W., {Jester} S., {Hennawi} J.~F.,
  {Dobler} G., 2010, \apj, 714, 1194

\bibitem[{{Schmidt} {et~al}\mbox{.}(2012){Schmidt}, {Rix}, {Shields}, {Knecht},
  {Hogg}, {Maoz}, \& {Bovy}}]{sch12}
{Schmidt} K.~B., {Rix} H.-W., {Shields} J.~C., {Knecht} M., {Hogg} D.~W.,
  {Maoz} D., {Bovy} J., 2012, \apj, 744, 147

\bibitem[{{Schneider} {et~al}\mbox{.}(2010){Schneider}, {Richards}, {Hall},
  {Strauss}, {Anderson}, {Boroson}, {Ross}, {Shen}, {Brandt}, {Fan}, {Inada},
  {Jester}, {Knapp}, {Krawczyk}, {Thakar}, {Vanden Berk}, {Voges}, {Yanny},
  {York}, {Bahcall}, {Bizyaev}, {Blanton}, {Brewington}, {Brinkmann},
  {Eisenstein}, {Frieman}, {Fukugita}, {Gray}, {Gunn}, {Hibon}, {Ivezi{\'c}},
  {Kent}, {Kron}, {Lee}, {Lupton}, {Malanushenko}, {Malanushenko}, {Oravetz},
  {Pan}, {Pier}, {Price}, {Saxe}, {Schlegel}, {Simmons}, {Snedden}, {SubbaRao},
  {Szalay}, \& {Weinberg}}]{sch10}
{Schneider} D.~P. {et~al.}, 2010, \aj, 139, 2360

\bibitem[{{Sergeev} {et~al}\mbox{.}(2005){Sergeev}, {Doroshenko},
  {Golubinskiy}, {Merkulova}, \& {Sergeeva}}]{ser05}
{Sergeev} S.~G., {Doroshenko} V.~T., {Golubinskiy} Y.~V., {Merkulova} N.~I.,
  {Sergeeva} E.~A., 2005, \apj, 622, 129

\bibitem[{{Sesar} {et~al}\mbox{.}(2007){Sesar}, {Ivezi{\'c}}, {Lupton},
  {Juri{\'c}}, {Gunn}, {Knapp}, {DeLee}, {Smith}, {Miknaitis}, {Lin}, {Tucker},
  {Doi}, {Tanaka}, {Fukugita}, {Holtzman}, {Kent}, {Yanny}, {Schlegel},
  {Finkbeiner}, {Padmanabhan}, {Rockosi}, {Bond}, {Lee}, {Stoughton}, {Jester},
  {Harris}, {Harding}, {Brinkmann}, {Schneider}, {York}, {Richmond}, \& {Vanden
  Berk}}]{ses07}
{Sesar} B. {et~al.}, 2007, \aj, 134, 2236

\bibitem[{{Shakura} \& {Sunyaev}(1973)}]{sha73}
{Shakura} N.~I., {Sunyaev} R.~A., 1973, \aap, 24, 337

\bibitem[{{Shappee} {et~al}\mbox{.}(2014){Shappee}, {Prieto}, {Grupe},
  {Kochanek}, {Stanek}, {De Rosa}, {Mathur}, {Zu}, {Peterson}, {Pogge},
  {Komossa}, {Im}, {Jencson}, {Holoien}, {Basu}, {Beacom}, {Szczygie{\l}},
  {Brimacombe}, {Adams}, {Campillay}, {Choi}, {Contreras}, {Dietrich},
  {Dubberley}, {Elphick}, {Foale}, {Giustini}, {Gonzalez}, {Hawkins}, {Howell},
  {Hsiao}, {Koss}, {Leighly}, {Morrell}, {Mudd}, {Mullins}, {Nugent},
  {Parrent}, {Phillips}, {Pojmanski}, {Rosing}, {Ross}, {Sand}, {Terndrup},
  {Valenti}, {Walker}, \& {Yoon}}]{sha14}
{Shappee} B.~J. {et~al.}, 2014, \apj, 788, 48

\bibitem[{{Shen} {et~al}\mbox{.}(2011){Shen}, {Richards}, {Strauss}, {Hall},
  {Schneider}, {Snedden}, {Bizyaev}, {Brewington}, {Malanushenko},
  {Malanushenko}, {Oravetz}, {Pan}, \& {Simmons}}]{she11}
{Shen} Y. {et~al.}, 2011, \apjs, 194, 45

\bibitem[{{Suganuma} {et~al}\mbox{.}(2006){Suganuma}, {Yoshii}, {Kobayashi},
  {Minezaki}, {Enya}, {Tomita}, {Aoki}, {Koshida}, \& {Peterson}}]{sug06}
{Suganuma} M. {et~al.}, 2006, \apj, 639, 46

\bibitem[{{Sun} {et~al}\mbox{.}(2014){Sun}, {Wang}, {Chen}, \& {Zheng}}]{sun14}
{Sun} Y.-H., {Wang} J.-X., {Chen} X.-Y., {Zheng} Z.-Y., 2014, \apj, 792, 54

\bibitem[{{Terlevich} {et~al}\mbox{.}(1992){Terlevich}, {Tenorio-Tagle},
  {Franco}, \& {Melnick}}]{tel92}
{Terlevich} R., {Tenorio-Tagle} G., {Franco} J., {Melnick} J., 1992, \mnras,
  255, 713

\bibitem[{{Tomita} {et~al}\mbox{.}(2006){Tomita}, {Yoshii}, {Kobayashi},
  {Minezaki}, {Enya}, {Suganuma}, {Aoki}, {Koshida}, \& {Yamauchi}}]{tom06}
{Tomita} H. {et~al.}, 2006, \apjl, 652, L13

\bibitem[{{Torricelli-Ciamponi} {et~al}\mbox{.}(2000){Torricelli-Ciamponi},
  {Foellmi}, {Courvoisier}, \& {Paltani}}]{tor00}
{Torricelli-Ciamponi} G., {Foellmi} C., {Courvoisier} T.~J.-L., {Paltani} S.,
  2000, \aap, 358, 57

\bibitem[{{Ulrich}, {Maraschi} \& {Urry}(1997){Ulrich}, {Maraschi}, \&
  {Urry}}]{ulr97}
{Ulrich} M.-H., {Maraschi} L., {Urry} C.~M., 1997, \araa, 35, 445

\bibitem[{{Uttley}(2006)}]{utt06}
{Uttley} P., 2006, in Astronomical Society of the Pacific Conference Series,
  Vol. 360, Astronomical Society of the Pacific Conference Series, {Gaskell}
  C.~M., {McHardy} I.~M., {Peterson} B.~M., {Sergeev} S.~G., eds., p. 101

\bibitem[{{Uttley} \& {Casella}(2014)}]{utt14}
{Uttley} P., {Casella} P., 2014, \ssr, 183, 453

\bibitem[{{Vanden Berk} {et~al}\mbox{.}(2001){Vanden Berk}, {Richards},
  {Bauer}, {Strauss}, {Schneider}, {Heckman}, {York}, {Hall}, {Fan}, {Knapp},
  {Anderson}, {Annis}, {Bahcall}, {Bernardi}, {Briggs}, {Brinkmann}, {Brunner},
  {Burles}, {Carey}, {Castander}, {Connolly}, {Crocker}, {Csabai}, {Doi},
  {Finkbeiner}, {Friedman}, {Frieman}, {Fukugita}, {Gunn}, {Hennessy},
  {Ivezi{\'c}}, {Kent}, {Kunszt}, {Lamb}, {Leger}, {Long}, {Loveday}, {Lupton},
  {Meiksin}, {Merelli}, {Munn}, {Newberg}, {Newcomb}, {Nichol}, {Owen}, {Pier},
  {Pope}, {Rockosi}, {Schlegel}, {Siegmund}, {Smee}, {Snir}, {Stoughton},
  {Stubbs}, {SubbaRao}, {Szalay}, {Szokoly}, {Tremonti}, {Uomoto}, {Waddell},
  {Yanny}, \& {Zheng}}]{van01}
{Vanden Berk} D.~E. {et~al.}, 2001, \aj, 122, 549

\bibitem[{{Vanden Berk} {et~al}\mbox{.}(2004){Vanden Berk}, {Wilhite}, {Kron},
  {Anderson}, {Brunner}, {Hall}, {Ivezi{\'c}}, {Richards}, {Schneider}, {York},
  {Brinkmann}, {Lamb}, {Nichol}, \& {Schlegel}}]{van04}
{Vanden Berk} D.~E. {et~al.}, 2004, \apj, 601, 692

\bibitem[{{Vaughan} {et~al}\mbox{.}(2003){Vaughan}, {Edelson}, {Warwick}, \&
  {Uttley}}]{vau03}
{Vaughan} S., {Edelson} R., {Warwick} R.~S., {Uttley} P., 2003, \mnras, 345,
  1271

\bibitem[{{Voevodkin}(2011)}]{voe11}
{Voevodkin} A., 2011, ArXiv e-prints: 1107.4244

\bibitem[{{Wall} \& {Jenkins}(2012)}]{wal12}
{Wall} J.~V., {Jenkins} C.~R., 2012, {Practical Statistics for Astronomers}

\bibitem[{{Wilhite} {et~al}\mbox{.}(2008){Wilhite}, {Brunner}, {Grier},
  {Schneider}, \& {vanden Berk}}]{wil08}
{Wilhite} B.~C., {Brunner} R.~J., {Grier} C.~J., {Schneider} D.~P., {vanden
  Berk} D.~E., 2008, \mnras, 383, 1232

\bibitem[{{Wilhite} {et~al}\mbox{.}(2005){Wilhite}, {Vanden Berk}, {Kron},
  {Schneider}, {Pereyra}, {Brunner}, {Richards}, \& {Brinkmann}}]{wil05}
{Wilhite} B.~C., {Vanden Berk} D.~E., {Kron} R.~G., {Schneider} D.~P.,
  {Pereyra} N., {Brunner} R.~J., {Richards} G.~T., {Brinkmann} J.~V., 2005,
  \apj, 633, 638

\bibitem[{{Winkler}(1997)}]{win97}
{Winkler} H., 1997, \mnras, 292, 273

\bibitem[{{Winkler} {et~al}\mbox{.}(1992){Winkler}, {Glass}, {van Wyk},
  {Marang}, {Jones}, {Buckley}, \& {Sekiguchi}}]{win92}
{Winkler} H., {Glass} I.~S., {van Wyk} F., {Marang} F., {Jones} J.~H.~S.,
  {Buckley} D.~A.~H., {Sekiguchi} K., 1992, \mnras, 257, 659

\bibitem[{{York} {et~al}\mbox{.}(2000){York}, {Adelman}, {Anderson},
  {Anderson}, {Annis}, {Bahcall}, {Bakken}, {Barkhouser}, {Bastian}, {Berman},
  {Boroski}, {Bracker}, {Briegel}, {Briggs}, {Brinkmann}, {Brunner}, {Burles},
  {Carey}, {Carr}, {Castander}, {Chen}, {Colestock}, {Connolly}, {Crocker},
  {Csabai}, {Czarapata}, {Davis}, {Doi}, {Dombeck}, {Eisenstein}, {Ellman},
  {Elms}, {Evans}, {Fan}, {Federwitz}, {Fiscelli}, {Friedman}, {Frieman},
  {Fukugita}, {Gillespie}, {Gunn}, {Gurbani}, {de Haas}, {Haldeman}, {Harris},
  {Hayes}, {Heckman}, {Hennessy}, {Hindsley}, {Holm}, {Holmgren}, {Huang},
  {Hull}, {Husby}, {Ichikawa}, {Ichikawa}, {Ivezi{\'c}}, {Kent}, {Kim},
  {Kinney}, {Klaene}, {Kleinman}, {Kleinman}, {Knapp}, {Korienek}, {Kron},
  {Kunszt}, {Lamb}, {Lee}, {Leger}, {Limmongkol}, {Lindenmeyer}, {Long},
  {Loomis}, {Loveday}, {Lucinio}, {Lupton}, {MacKinnon}, {Mannery}, {Mantsch},
  {Margon}, {McGehee}, {McKay}, {Meiksin}, {Merelli}, {Monet}, {Munn},
  {Narayanan}, {Nash}, {Neilsen}, {Neswold}, {Newberg}, {Nichol}, {Nicinski},
  {Nonino}, {Okada}, {Okamura}, {Ostriker}, {Owen}, {Pauls}, {Peoples},
  {Peterson}, {Petravick}, {Pier}, {Pope}, {Pordes}, {Prosapio},
  {Rechenmacher}, {Quinn}, {Richards}, {Richmond}, {Rivetta}, {Rockosi},
  {Ruthmansdorfer}, {Sandford}, {Schlegel}, {Schneider}, {Sekiguchi}, {Sergey},
  {Shimasaku}, {Siegmund}, {Smee}, {Smith}, {Snedden}, {Stone}, {Stoughton},
  {Strauss}, {Stubbs}, {SubbaRao}, {Szalay}, {Szapudi}, {Szokoly}, {Thakar},
  {Tremonti}, {Tucker}, {Uomoto}, {Vanden Berk}, {Vogeley}, {Waddell}, {Wang},
  {Watanabe}, {Weinberg}, {Yanny}, {Yasuda}, \& {SDSS Collaboration}}]{yor00}
{York} D.~G. {et~al.}, 2000, \aj, 120, 1579

\bibitem[{{Zheng} {et~al}\mbox{.}(1997){Zheng}, {Kriss}, {Telfer}, {Grimes}, \&
  {Davidsen}}]{zhe97}
{Zheng} W., {Kriss} G.~A., {Telfer} R.~C., {Grimes} J.~P., {Davidsen} A.~F.,
  1997, \apj, 475, 469

\bibitem[{{Zu} {et~al}\mbox{.}(2013){Zu}, {Kochanek}, {Koz{\l}owski}, \&
  {Udalski}}]{zu13}
{Zu} Y., {Kochanek} C.~S., {Koz{\l}owski} S., {Udalski} A., 2013, \apj, 765,
  106

\bibitem[{{Zuo} {et~al}\mbox{.}(2012){Zuo}, {Wu}, {Liu}, \& {Jiao}}]{zuo12}
{Zuo} W., {Wu} X.-B., {Liu} Y.-Q., {Jiao} C.-L., 2012, \apj, 758, 104

\end{thebibliography}

\label{lastpage}

\end{document}